\documentclass{article}[16pt]
\usepackage[utf8]{inputenc}
\usepackage{graphicx}
\begin{document}
\begin{center}

{\bf \Large SURFACE PHOTOMETRY OF 50 DWARF GALAXIES IN THE LOCAL VOLUME}\\ 

\bigskip

{\bf K.~A.~Kryzhanovsky$^1$, M.~E.~Sharina$^2$, I~D.~Karachentsev$^2$, and G.~M.~Karataeva$^1$}

\bigskip

{\bf $^1$  St. Petersburg State University, St. Petersburg, Russia} \\
{\bf $^2$  Special Astrophysical Obseratory of the Russian Academy of Sciences, Nizhnyi Arkhiz, Russia\\ 
e-mail: sme@sao.ru}
 
 \end{center}
  
\bigskip

{\bf Abstract}

The results of surface photometry of 50 galaxies in the Local Volume based on archival images obtained
with the Hubble Space Telescope are presented. For the sample of galaxies, the integrated magnitudes in the
V and I bands are given, as well as the brightness and color profiles. The obtained photometric parameters are
compared with the measurements of other authors.


Original article submitted April 28, 2023.  Translated from Astrofizika, Vol. 66, No. 3, pp. 317-329 (August 2023)

\section{Introduction.}

 The standard cosmological model $\Lambda$CDM successfully explains the
major properties of the large scale structure of the Universe. On going to smaller scales on the
order of 1 Mpc, some well known discrepancies between the theoretical predictions and
observational data show up. The most appropriate target for comparison of the results of the
theory with observations is the Local Volume (LV) with a radius of 10--12~Mpc around the Milky
Way, for which the density of observational data is greater than for distant regions. This radius
has been chosen because the distance of each galaxy inside it can be measured with an accuracy
of about 5\% on the Hubble Space Telescope, all within a single orbital period. The first list of
galaxies in the Local Volume contained just 179 objects [1]. Targeted searches of the close
galaxies led to the creation of ``A Catalog of Neighboring Galaxies''[2] with $N = 450$ members
and an ``Updated Nearby Galaxy Catalog'' [3] with $N = 869$. Over recent years digital surveys of
large segments of the sky in the optical range and in the hydrogen 21-cm line (SDSS, DECaLS,
HIPASS, ALFALFA, etc.) have greatly increased the population of the galaxies in the LV. The
latest version of the Local Volume Galaxy Database ([4];  www.sao.ru/lv/lvgdb) contains more than 1300
objects. A large fraction of these have accurate estimates of distances and radial velocities,
which is particularly important for analyzing the distribution of dark matter in the LV. The
integrated luminosity of a galaxy, along with its radial velocity and distance, is one of the most
important parameters of the galaxy. Because of the rapid growth in the population of the LV,
many close dwarf galaxies with a low surface brightness ended up without reliable photometry,
with just crude visual estimates of the visible magnitudes. A need for significant strengthening
of the photometric base for the galaxies of the LV has become evident.

The main purpose of this paper is to obtain photometric parameters of the galaxies with
subsequent analysis of the results in the form of a comparison of them with the values taken from
other publications, as well as surveys and catalogs. In this paper we do not study the properties
of the galaxies, or make detailed interpretation of the results and construct various dependences,
but only provide for the possibility of using these results in later papers. Observations of the
galaxies were made in 2019--2020 using the Advanced Camera for Surveys (ACS) installed on
the Hubble Space Telescope (HST) in the F606W and F814W filters as part of the SNAP 15922
project (``Every Known Nearby Galaxy'', PI R. B. Tully). The observations yielded color-
magnitude diagrams for the stellar population of 80 galaxies. Of these, the distances with respect
to the luminosity of the tip of the red giants branch were determined for 63 of the galaxies.
Images of the observed galaxies, color-magnitude diagrams, and the measured distances are
given in the ``Extragalactic Distance Database'' (EDD [5]) with supplements in Ref. 6.

\section{Photometry.} The surface photometry process was analogous to that in Refs. 7 and 8.
The SURFPHOT program package was used to carry out the photometry. This is part of a large
packet of programs $MIDAS$ (Munich Image Data Analysis System) [9] for analyzing
astronomical data developed at the ESO. A corresponding script was written for carrying out the
different commands and programs of this package. Circular and elliptical apertures were used to
obtain the integrated magnitudes. The procedures for searching for the galactic centers and
modeling the distribution of the intensity over the area of an object were carried out with the aid
of the procedure $FIT/ELL3$ for inscribing ellipses. The sky background was evaluated and
subtracted from the image using the procedure $FIT/FLAT SKY$ which creates a two-dimensional
polynomial by a method of least squares. Background objects were resolved with the aid of
$SExtractor$ 2.5.0 ([10,11]). The flux was integrated in the obtained apertures (procedures
$INTEGRATE/APERTURE$ and $INTEGRATE/ELLIPS$) and the azimuthally averaged surface
brightnesses were calculated.
The photometric results were converted into the standard Johnson-Cousins BVRI system
with the aid of the empirical formulas
\begin{eqnarray}
                        V = V_i + 0.236 (V_i - I_i) + 26.325;   
                        (V - I) = 1.309 (V_i - I_i) + 0.83; 
                        I = I_i + 25.495,
\end{eqnarray}

where the quantities with subscript $i$ are measured in the instrumental photometric system (see
Ref. 12). The resulting integrated stellar magnitudes were used to calculate the $B$ band
magnitudes using the formula [13]
\begin{eqnarray}
  \rm B' = V +0.85(V - I) - 0.20.
  \end{eqnarray}

The parameters of the profiles for the galaxies were obtained using a model based on the
exponential function [14]: $ \rm \mu(r) = \mu_0 +1.086(r/h),$
 $\mu_0$, where is the central surface brightness and $h$~--- is an
exponential scale length.

\section{Results.} 
The main results are presented in Table 1.\footnote
{Other detailed data from our photometry are available upon individual request from the contact
author of this article.}

Table 1 contains a list of the photometrized galaxies named in columns 1 and 2 with the
coordinates at epoch J2000.0 given in column 3. The morphological types of the galaxies and
the published [3] distances to them in Mpc are given in columns 4 and 5. Columns 6, 7, and 8
contain the results of this photometry and an analysis of the brightness profiles in this paper: the
central surface brightnesses in the $V$-band and the integrated stellar magnitudes in the V
and $I$-bands of the Johnson-Cousins photometric system. Column 9 lists the stellar magnitudes in
the $B$-band of the Johnson-Cousins system calculated from the measured $V$ and $I$ stellar
magnitudes using Eq. (2). None of the stellar magnitudes given in Table 1 are corrected for
absorption. Column 10 contains the stellar magnitudes of the galaxies in the $B$-band of the
Johnson-Cousins system from the literature. Column 11 lists the values for the absorption of
light in the galaxy in the B band in stellar magnitudes. Published references to the data shown in
column 10 are listed in column 12. ``LV'' indicates a reference to the latest version of the data
base of Ref. 4 (and references in it), or to individual visual estimates of the stellar magnitude by
I. D. Karachentsev.
The profiles of the surface brightness obtained by photometry are shown in Fig. 1. \footnote{ 
Figure 1 is continued at the end of this article.} %Because of the small size of the frames in Fig. 1,
%the error bars are not clearly visible.}

The following frames are shown for each galaxy: top left- plots of the integrated stellar magnitude in
the $V$ (dark curve) and $I$ (light curve) bands; bottom left- the corresponding difference between
the $V$ and $I$ growth curves; top right- the $V$ and $I$ brightness curves in mag./arcsec$^2$; bottom
right, the corresponding difference between the $V$ and $I$ brightness profiles. The error range in
the photometry is indicated by light bars. For more detail on determining the photometric errors,
see Ref. 15.

A comparison of the results of the photometry converted into B'; magnitudes (Eq. (2))
with literature photometric estimates (Table 1, column 10) yields an average difference $\rm \Delta B = <B' - B_{lit}> = +0.06\pm 0.07^m$ and
a standard error of $\rm \sigma(\Delta B) =0.30^m$. After the photometric errors in our data and the published data are taken
into account, the error in our measurements is 0.2$^m$.

\section{Concluding comments.} This paper presents the results of surface photometry of 50
galaxies in the Local Volume with distances $D<12$~Mpc. The photometry is based on images of
the galaxies obtained on the Hubble Space Telescope in the $V$ and $I$ bands as part of the SNAP
15922 program. The objects chosen for measurement had diameters that did not exceed he
angular dimensions of the ACS camera of the HST. The integrated $V$ and $I$ magnitudes of the
galaxies were determined and profiles of the surface brightness were constructed in both bands.
A comparison of the resulting integrated magnitudes of the galaxies with data available from
other sources shows that the error in our estimates of the integrated magnitudes is about 0.2$^m$.
Most of the galaxies studied here are objects with low surface brightness at a median magnitude
of $SB_{\rm OV}\simeq 22.2$~mag./arcsec$^2$ .

The authors thank L. N. Makarova and D. I. Makarov for useful discussions. This work
is based on observations made with the NASA/ESA Hubble Space Telescope and obtained from
the Hubble Legacy Archive created in a collaboration between the Space Telescope Science
Institute (STScI/NASA), the Space Telescope European Coordinating Facility (ST-
ECF/ESAC/ESA) and the Canadian Astronomy Data Centre (CADC/NRC/CSA). The SNAP-
15922 program (P.I., R. B. Tully) was supported by NASA through a grant from the Space
Telescope Science Institute under the direction of the Association of Research Universities in the
area of astronomy under the contract NASA NASb5-26555.
This work was partially supported by the Ministry of Science and Higher Education of
the Russian Federation, grant No. 075--15--2022--262 (13.MNPMU.21.0003).

\bigskip

{\bf REFERENCES}

1. R. C. Kraan-Korteweg, G. A. Tammann, Astron. Nachr., 300, 181 (1979).

2. I.D. Karachentsev, V.E. Karachentseva, W.K. Huchtmeier, D.I. Makarov, Astron. J., 127, 2031 (2004).

3. I.D. Karachentsev, Makarov D.I., Kaisina E.I., Astron. J., 145, 101 (2013).

4. E. I. Kaisina, D. I. Makarov, I. D. Karachentsev, S. S. Kaisin, Astrophys. Bull., 67, 115 (2012).

5. G. S. Anand, L. Rizzi, Tully R.B., Astron. J., 162, 80 (2021).

6. I.D. Karachentsev and N. A. Tikhonov, Astrophysics, 66, 1 (2023).

7. Swaters, R. A. and Balcells, M., A\&A 390 , 863 (2002).

8. Taylor V. A., Jansen R. A, Windhorst R. A., et al., Astrophys. Journal 630, 784 (2005).

9. Banse K., Crane P., Grosbol P., et al., The Messenger 31, 26 (1983).

10. D. O. Cook, D. A. Dale, B. D. Johnson, et al., Mon. Not. Roy. Astron. Soc., 445, 881 (2014).

11. L. C. Ho, Z.-Y. Li, A. J. Barth, et al., Astrophys. J. Suppl. Ser. 197, 21 (2011).
12. M. Sirianni, M. J. Jee, N. Benitez, et al., Publ. Astron. Soc. Pacific 117 (836), 1049 (2005).

13. Makarova L., Astron. and Astrophys., 139, 491 (1999).

14. de Vaucouleurs G., 1959, in Flugge S., ed., Handbuch der Physik 53. Springer-Verlag, Berlin, p. 275

15. Sharina M. E. et al., Mon. Not. Roy. Astron. Soc., 384, 1544 (2008).

16. Maddox S. J., Efstathiou G., Sutherland W. J., Mon. Not. Roy. Astron. Soc., 246, 433 (1990).

17. Doyle, M. T.; Drinkwater, M. J.; Rohde, D. J. et al., Mon. Not. Roy. Astron. Soc., 361, 34 (2005).

18. L.N. Makarova, I.D. Karachentsev, E.K. Grebel, O.Yu. Barsunova, Astron. and Astrophys., 384, 72
(2002).

19. L.N. Makarova, Astron. and Astrophys. Suppl., 128, 459 (1998).

20. Jones D. H., Saunders W., Read, M., Colless M., Publications of the Astron. Society of Australia, 22,
277 (2005).

21. Colless M., Peterson B.A., Jackson C., et al., eprint arXiv:astro-ph/0306581 (2003).

22. Sharina M. E., Astrophysics, 62, 9 (2019).

23. I. D. Karachentsev, L. N. Makarova, B. S. Koribalski, G. S. Anand, R. B. Tully, A. Y. Kniazev, Mon.
Not. Roy. Astron. Soc., 518, 5893 (2023).

24. S. Alam, F. D. Albareti, C. A. Prieto, et al., Astrophys. Journal Suppl. Ser., 219, 12 (2015).

25. J. K. Adelman-McCarthy, M. A. Agueros, S. S. Allam, et al., Astrophys. Journal Suppl. Ser., 175, 297
(2008).

26. Impey C. D., Sprayberry D., Irwin M. J., Bothun G. D., Astrophys. Journal Suppl. Ser., 105, 209 (1996).

27. H. Aihara and P. Mcgehee, Astrophys. Journal Supplement Series 193 (2011).

28. de Vaucouleurs G., de Vaucouleurs A., Corwin H. Jr, Buta R. J., Paturel G., Fouqu P., 1991, Third
Reference Catalogue of Bright Galaxies. Springer-Verlag, New York

29. A. Gil de Paz, B. F. Madore, and O. Pevunova, Astrophys. J. Suppl. 147, 29 (2003).

30. J. Koda, M. Yagi, Y. Komiyama, et al., Astrophys. J. 802, L24 (2015).

\clearpage
\begin{table}
\caption{Surface photometry of 50 galaxies in the Local Volume observed with the HST in the SNAP 15922 survey}
 \label{tab:properties}
 \scriptsize
\begin{center}
\begin{tabular}{lccccccccccl}
\hline\hline
 Name         &     PGC   & RA (2000.0) DEC  &Type     &D     &SBv0     & V     & I      & B'    & B$_{lit}$& A$_B$& Ref. \\
 (1)          &      (2)  &     (3)          & (4)     & (5)  &   (6)   & (7)   & (8)    & (9)   & (10)    & (11) & (12)  \\
\hline                                                                                                                                  
 UGC 064      &    000591 &  000744.0+405232 &   dIrr  &  8.16 &  21.49 &  15.44 &  14.97 &  15.65 & 15.5   & 0.34& LV    \\
 WOC2017-07   &     -     &  005501.0-231009 &   dIrr  &  3.62 &  24.78 &  17.90 &  17.37 & 18.15  &  18.1  & 0.07& [16]  \\
 AGC 122226   &   086806  & 024638.9+274335  &  BCD    &  7.71 &  21.24 &  15.75 &  15.03 &  16.17 &  17.1  & 0.53& LV    \\
 ESO 300-016  &   011842  &  031010.5-400011 &   dIrr  &  9.33 &  22.04 &  15.92 &  15.52 &  16.06 &   15.6 & 0.08& LV    \\
 UGC 2716     &    012719 & 032407.2+174515  & Sm      &  6.66 &  21.47 &  14.07 &  13.36 &   14.47&   14.6 & 0.59& LV    \\
 KKH 22       &   2807114 &  034456.6+720352 &  dTr    &  3.12 &  25.00 &  16.22 &  15.04 &  17.02 &   18.0 & 1.66& LV    \\
 HIPASSJ0517  &  4078612  & 051721.6-324535  &  dIm    &  9.32 &  21.13 &  15.57 &  15.11 &  15.76 &   15.7 & 0.07& [17]  \\
 KKH 34       &   095594  &  055941.2+732539 &   dIrr  &  7.28 &  24.04 &  17.02 &  16.20 &  17.52 &   17.1 & 1.08& LV    \\
 ESO 006-001  &   023344  &  081923.3-850844 &   dTr   &  2.70 &  22.32 &  14.68 &  13.65 &  15.36 &     -  & 0.83&    -  \\
 KKH 46       &  2807128  & 090836.6+051732  & dIrr    &  6.70 &  23.65 &  16.45 &  15.99 &  16.65 &  17.0  & 0.20& [18]  \\
 ESO 373-007  &   027104  &  093245.4-331444 &   dIrr  &  9.77 &  23.48 &  15.15 &  14.32 &  15.66 &  16.4  & 0.58& LV    \\
 UGC 5086     &   027115  & 093248.9+212754  & dSph    &  8.49 &  22.67 &  15.97 &  15.16 &  16.44 &  15.9  & 0.14& [19]  \\
 6dFJ0944201  &   807172  &  094420.1-225458 &   BCD   &  10.47&  21.42 &  17.12 &  16.73 &  17.23 &  16.8  & 0.33& [20]  \\
 2MASXJ0957   &  154449   & 095708.9-091548  &  BCD    & 10.13 &  20.17 &  15.27 &  14.45 &  15.77 &  15.8  & 0.29& [20]  \\
 MCG -01-26   &   029033  & 100138.4-081456  &  dIrr   &  9.94 &  22.24 &  14.97 &  14.22 &  15.40 &  15.4  & 0.15& LV    \\
 2dFGRS-N21   & 1099440   & 100932.5-021058  & BCD     & 10.42 &  20.92 &  15.58 &  15.16 &  15.75 &  15.8  & 0.19& [21]  \\
 UGC 5918     &   032405  &  104936.5+653150 &  dIrr   &  8.50 &  23.68 &  14.81 &  14.06 &  15.25 &  15.0  & 0.05& [13]  \\ 
 Mrk 1265     &   032413  &  104940.4+225019 &  BCD    &  9.55 &  20.63 &  15.13 &  14.68 &  15.31 &  17.0  & 0.09 & LV   \\
 KKH 68       &   2807141 &  113053.3+140846 &  dIrr   & 12.47 &  23.29 &  16.12 &  15.37 &  16.56 &   16.6 & 0.17& [22]  \\
 HIPASSJ1131  &  5060432  & 113135.2-314020  &  dIrr   & 6.90  &  22.43 &  18.06 &  17.59 &  18.14 &  18.2  & 0.30& [23]  \\
 KKH 69       &   2807142 &  113453.3+110112 &  dIrr   & 7.40  &  23.53 &  16.80 &  16.13 &  17.17 &  16.6  & 0.10& [22]  \\
 LBTJ115205   &   [Grapes]&  115205.6+544732 &  dIrr   & 5.96  &  23.75 &  18.10 &  17.64 &  18.27 &  18.5  & 0.04& LV    \\
 EVCC 67      &    4304796&  115840.4+153534 &  dIrr   & 16.5  &  21.12 &  17.71 &  17.02 &  18.10 &  18.2  & 0.08& [24]  \\
 ESO 379-024  &   038252  &  120456.7-354435 &   dIrr  & 5.46  &  22.34 &  16.44 &  16.22 &   16.43&   16.6 & 0.33 & LV   \\
 SDSSJ1205    &   4310323 &  120531.0+310434 &  Sdm    & 16.0  &  21.49 &  17.20 &  16.34 &   17.73&   17.6 & 0.08&  [25] \\
 KK 135       &   166130  &  121934.7+580234 &  dIrr   & 5.46  &  24.03 &  17.50 &  17.33 &   17.45&   18.1 & 0.05& LV    \\
 MCG+09-20    & 040750    & 122652.6+530619  & BCD     & 6.12  &  21.67 &  15.74 &  15.25 &   15.95&   15.9 & 0.10& LV    \\
 MCG+00-32    & 041395    & 123103.8+014033  & dIm     & 9.42  &  21.81 &  15.40 &  14.31 &   16.13&   15.9 & 0.08&  [26] \\
 WSRT-CVN43   &     -     & 123109.0+420539  & dIrr    & 8.13  &  23.15 &  17.55 &  17.48 &   17.41&   17.8 & 0.08&  LV   \\
 MCG+07-26    &   041749  &  123352.7+393733 &  dIm    & 9.94  &  22.34 &  15.42 &  15.04 &   15.54&   16.5 & 0.06&  [27] \\
 KUG1234+29   &   042115  &  123714.0+293751 &  BCD    & 8.44  &  20.16 &  15.35 &  14.67 &   15.73&   16.3 & 0.07&  [24] \\
 KKSG 30      &  3097708  &  123735.9-085202 &  dIrr   & 9.73  &  23.55 &  16.17 &  15.56 &   16.47&   16.3 & 0.14&  LV   \\
 UGC 7827     &    042380 &  123938.9+444915 &  dIrr   & 9.09  &  22.73 &  15.11 &  14.78 &   15.19&   16.0 & 0.08& LV    \\
 KDG 178      &    042413 &  124010.0+323931 &  dIrr   & 13.0  &  22.80 &  16.06 &  15.61 &   16.22&   17.1 & 0.06& LV    \\
 SDSSJ1240    &   4074723 &  124029.9+472204 &  dIrr   & 7.63  &  23.93 &  17.97 &  17.72 &   17.97&   18.2 & 0.07& LV    \\
 NGC 4627     &    042620 &  124159.7+323425 &  E      & 6.93  &  20.32 &  12.74 &  11.88 &   13.27&   13.1 & 0.07& [28]  \\
 BTS 151      &   2832120 &  124324.6+322856 &  dSph   & 7.60  &  23.01 &  16.85 &  16.10 &   17.29&   17.6 & 0.07& [25]  \\
 UGC 7903     &    042832 &  124345.0+535732 &  dIrr   & 9.66  &  23.52 &  16.32 &  15.87 &   16.50&   16.6 & 0.06& LV    \\
 ESO 219-010  &   044110  &  125609.6-500838 &   dSph  & 4.29  &  22.51 &  15.32 &  14.19 &   16.08&   16.4 & 0.96 & LV   \\
 MCG+07-27    &  045889   &  131251.8+403235 &  BCD    & 8.99  &  20.42 &  14.24 &  13.58 &   14.60&   14.9 & 0.06&  LV   \\
 KK 195       &   166163  &  132108.2-313147 &   dIrr  & 5.62  &  23.63 &   17.03&   16.60&   17.20&   17.1 & 0.27& LV    \\
 NGC 5229     &    047788 &   133402.9+475455&   Sdm   & 8.95  &  21.09 &   13.51&   12.71&   13.99&   14.3 & 0.08& [13]  \\
 KKs 58       &   2815824 &  134600.8-361944 &   dSph  & 3.75  &  23.06 &   16.39&   15.55&   16.90&   17.4 & 0.27& LV    \\
 ESO 222-010  &   052125  &  143503.0-492518 &   dIrr  & 3.15  &  22.34 &  14.87 &   14.13&   15.31&   16.3 & 1.11& LV    \\
 Mrk 475      &   052358  &  143905.4+364822 &  BCD    &11.53  &  20.06 &  15.56 &  15.35 &   15.54&   16.3 & 0.05& [29]  \\
 ESO 272-025  &   052591  &  144325.5-444219 &   dIrr  & 3.91  &  21.28 &  13.88 &  13.05 &   14.39&   14.8 & 0.69 & LV   \\
 KK 242       &   4689184 &  175248.4+700814 &  dTr    & 6.46  &  24.99 &  18.38 &   17.40&   19.02&   18.6 & 0.14& [30]  \\
 AGC 322463   &  5067080  &  225935.3+164611 &  dIrr   & 7.97  &  22.05 &  17.41 &  16.95 &   17.60&   17.2 & 0.58& LV    \\
 ESO 347-017  &   071464  &  232656.1-372049 &   dIm   & 8.42  &  21.52 &  14.00 &  13.48 &   14.25&   14.7 & 0.07& LV    \\
 2DFGRS-S43   &   704814  &  235840.7-312803 &   dIrr  & 3.66  &  22.22 &  15.85 &  14.90 &   16.47&   16.2 & 0.07& [16]  \\
\hline\hline                                                                                                                                 
\end{tabular}
%Notes: $^a$: Karachentsev et al. (2015b), $^ b$: Karachentsev et al. (2013), 
%$^c$:~Karachentsev, Kniazev \& Sharina (2015a), $^d$: Makarova et al. (2007), $^e$ Sharina et al. (2008).
\end{center}
\end{table}

\clearpage
\begin{figure*}
\begin{minipage}[h]{0.47\linewidth}
\includegraphics[scale=0.27,angle=-90]{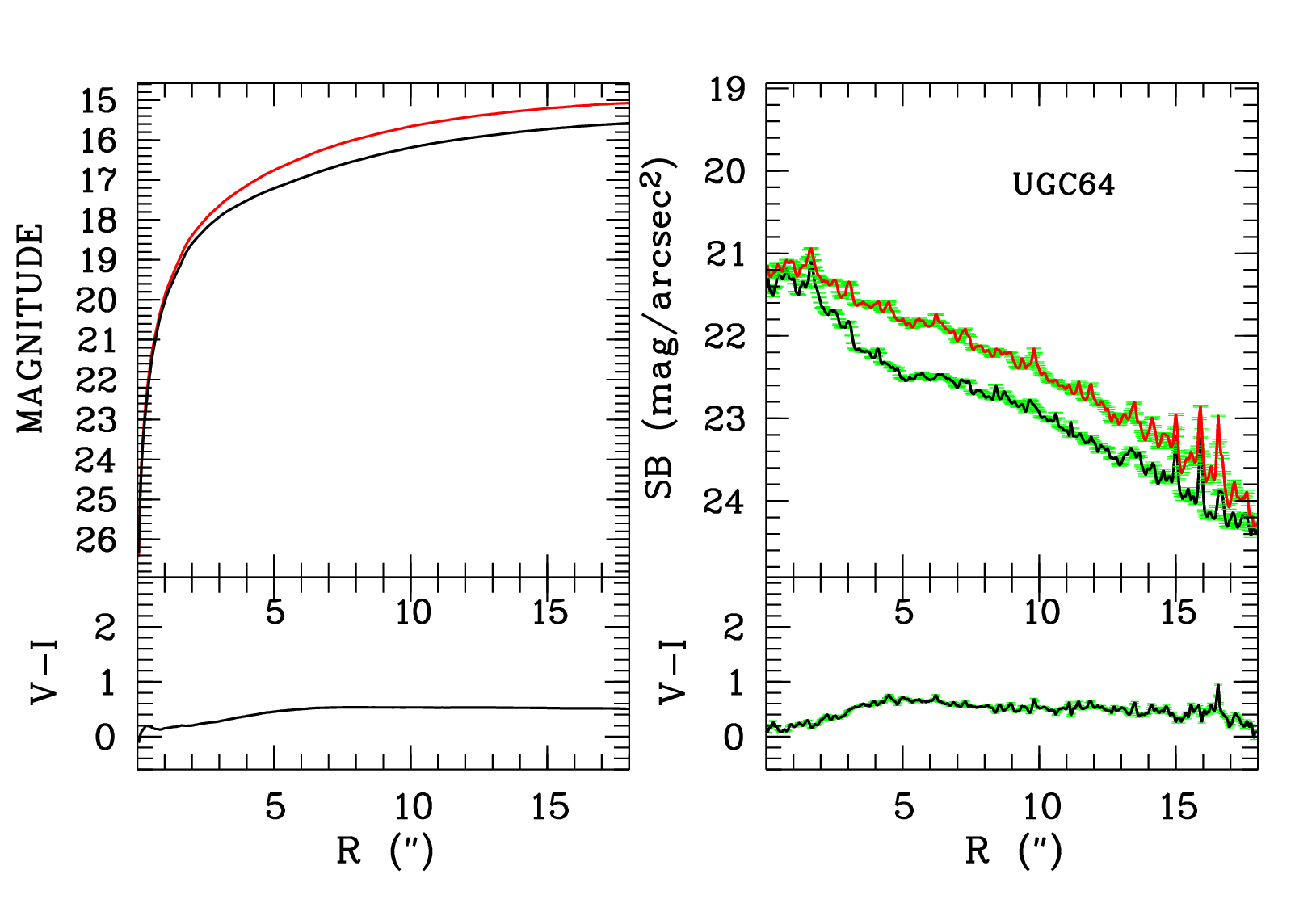}
\end{minipage}
\hfill
\begin{minipage}[h]{0.47\linewidth}
\includegraphics[scale=0.27,angle=-90]{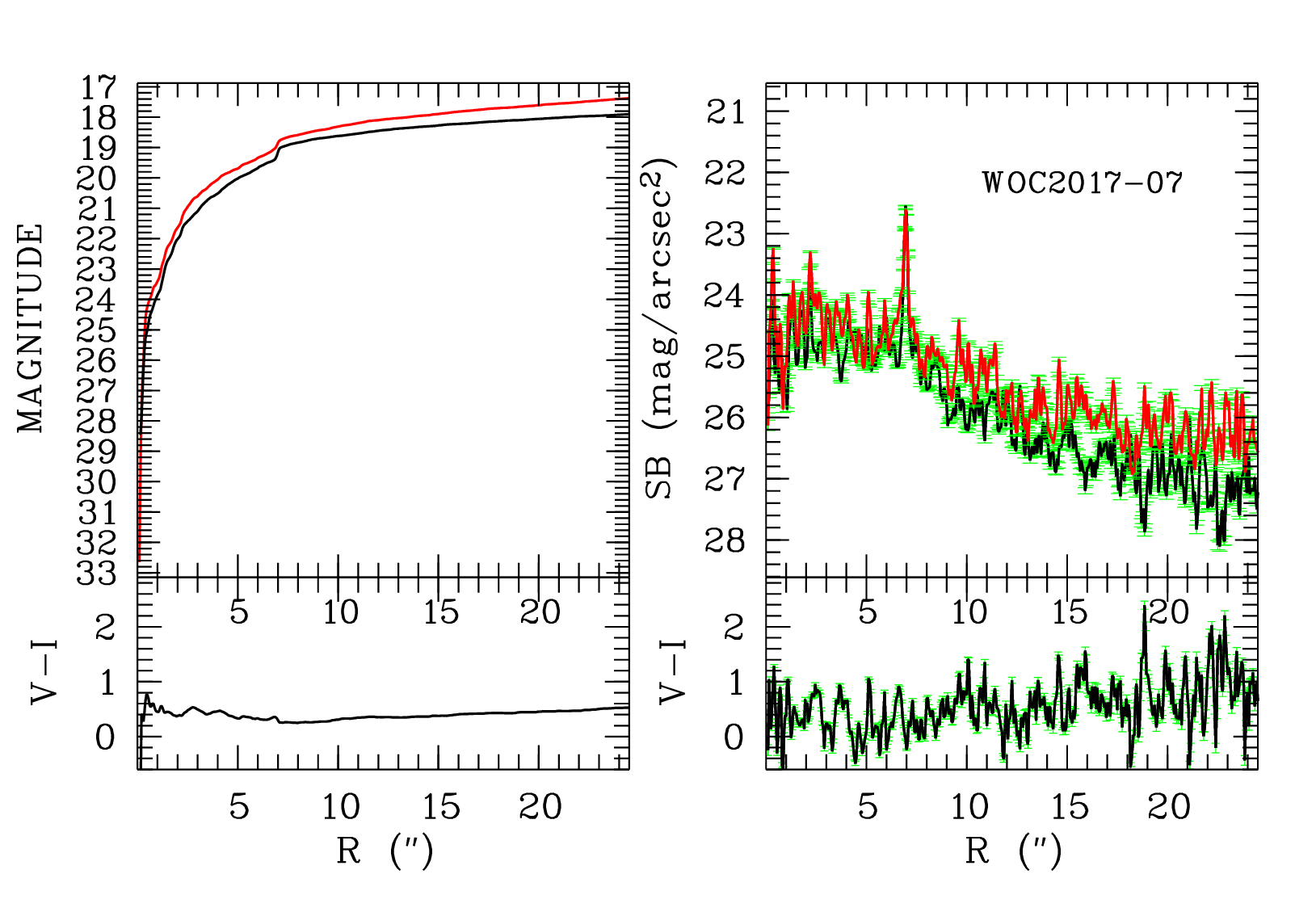}
\end{minipage}
\caption{Profiles of the surface brightness and the growth curves for the integrated stellar magnitude of the 50 galaxies.
}
\end{figure*}

\setcounter{figure}{0}
\begin{figure*}
\begin{minipage}[h]{0.47\linewidth}
\includegraphics[scale=0.27,angle=-90]{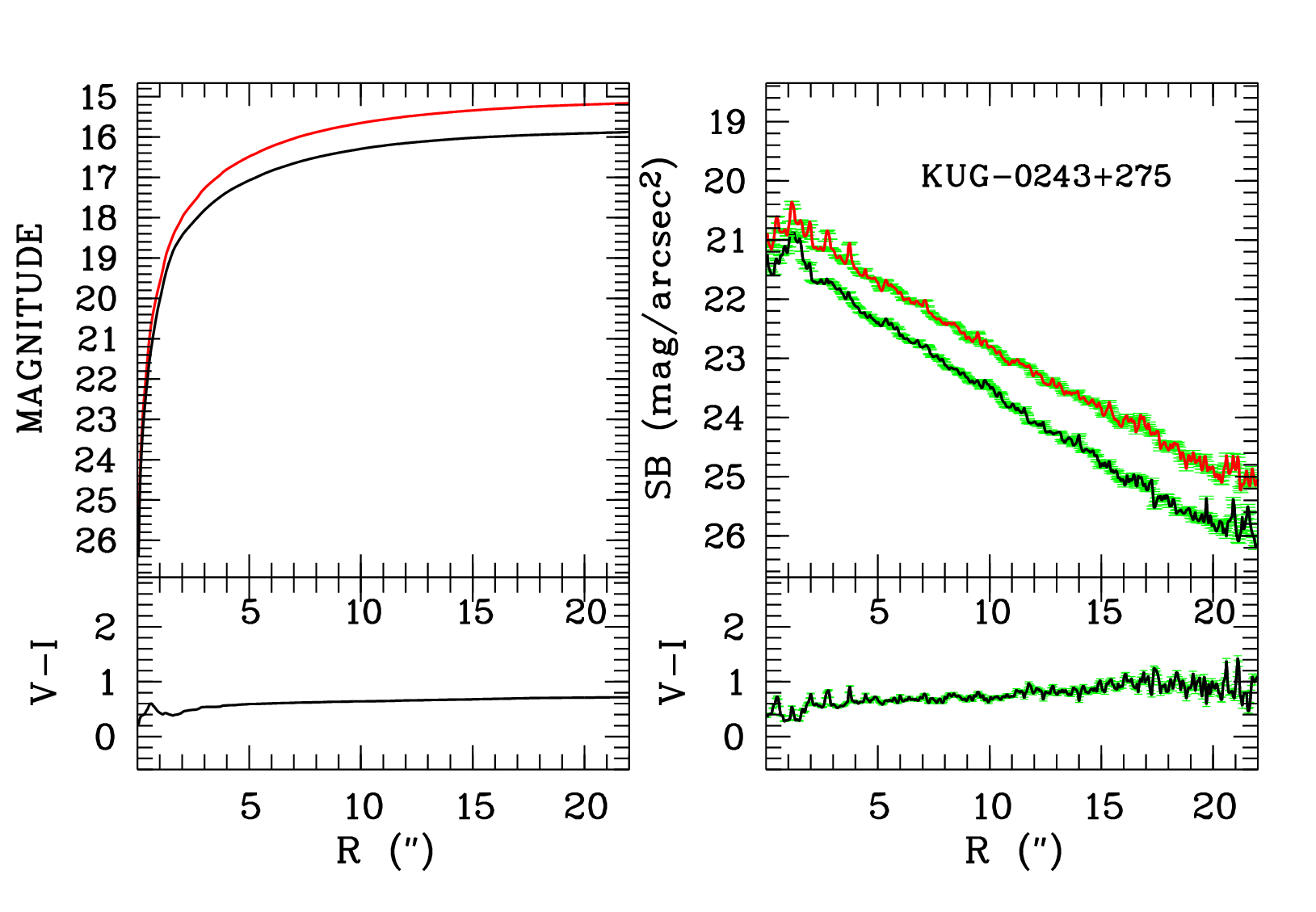}
\end{minipage}
\hfill
\begin{minipage}[h]{0.47\linewidth}
\includegraphics[scale=0.27,angle=-90]{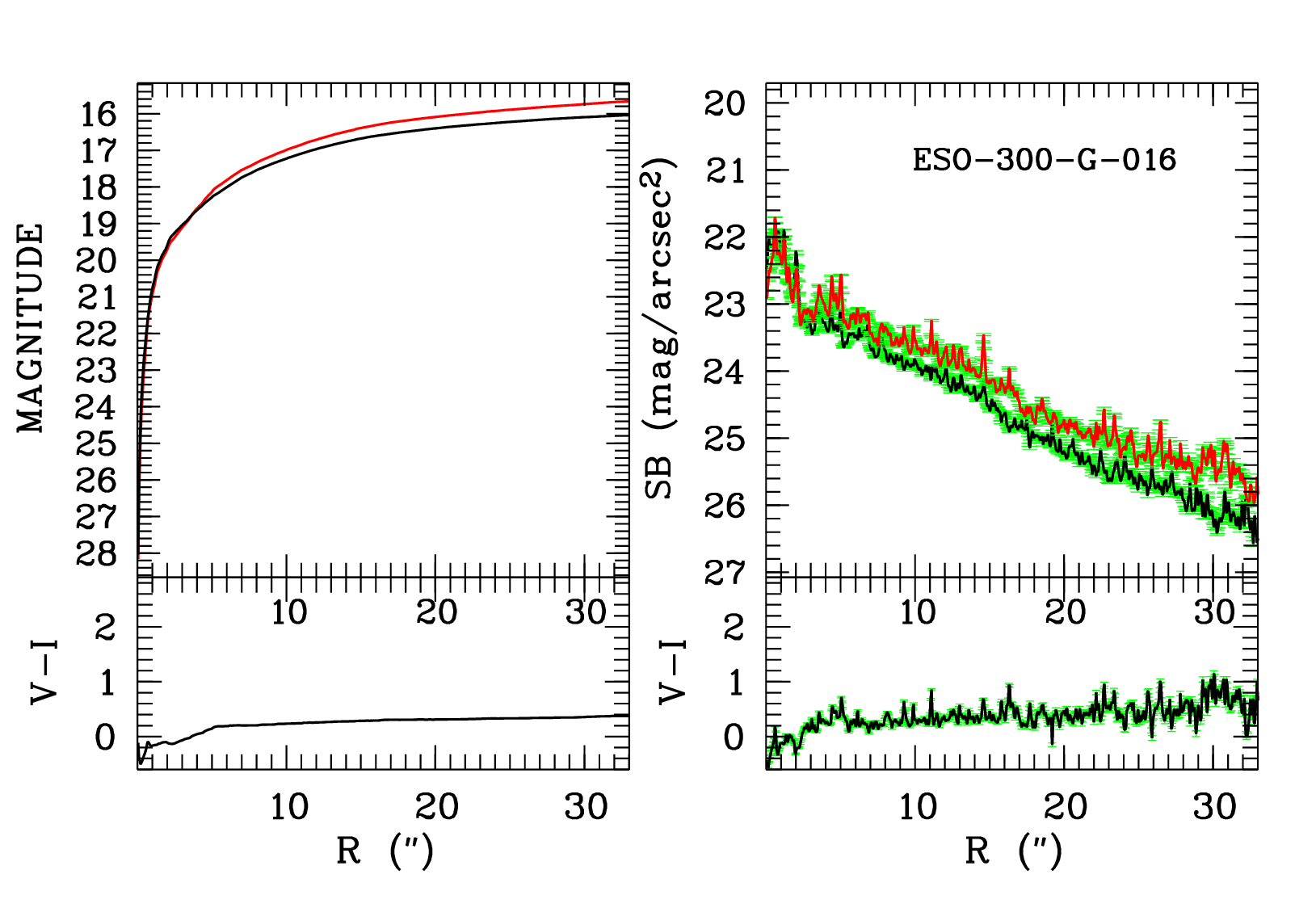}
\end{minipage}
\vfill
\begin{minipage}[h]{0.47\linewidth}
\includegraphics[scale=0.27,angle=-90]{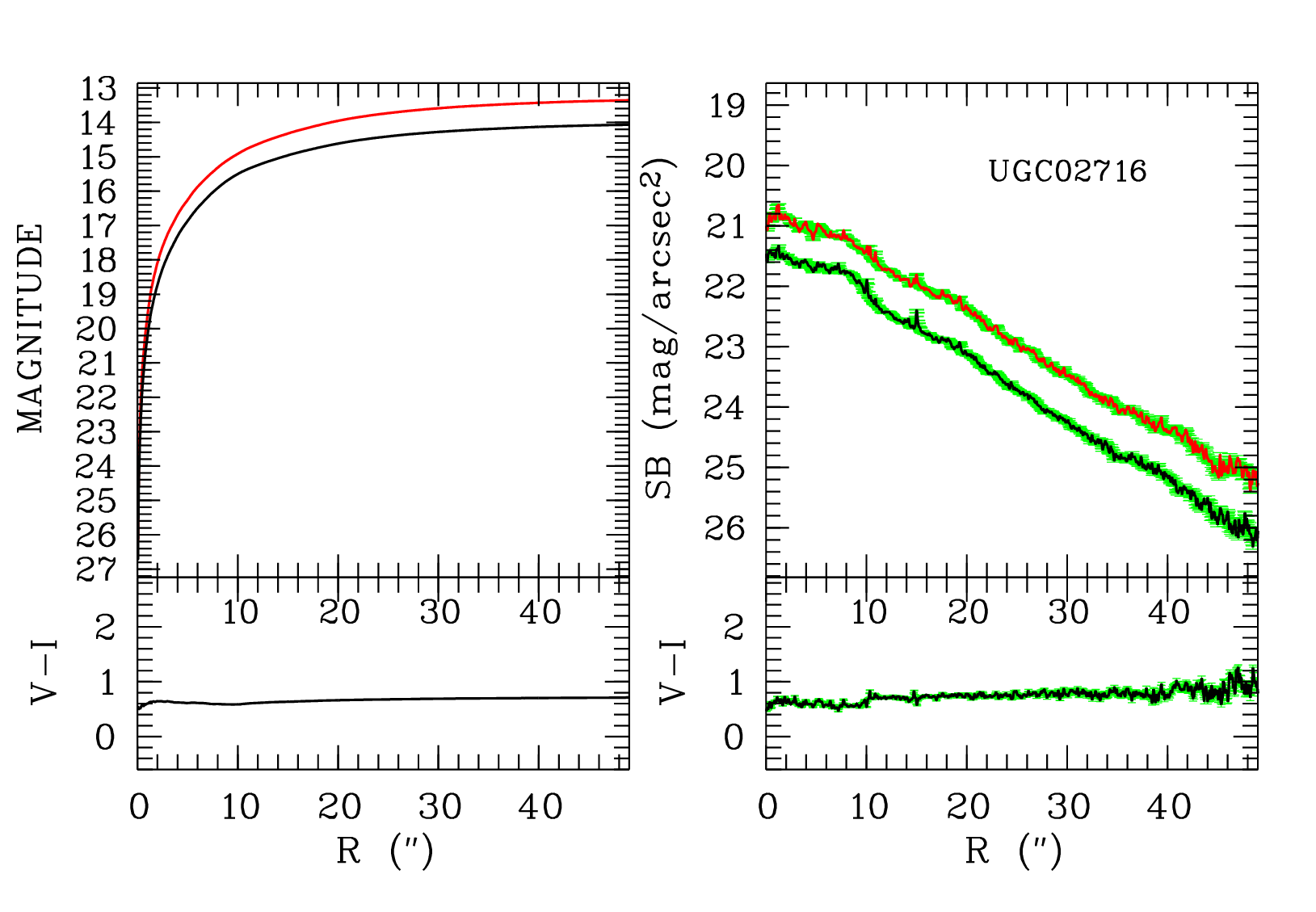}
\end{minipage}
\hfill
\begin{minipage}[h]{0.47\linewidth}
\includegraphics[scale=0.27,angle=-90]{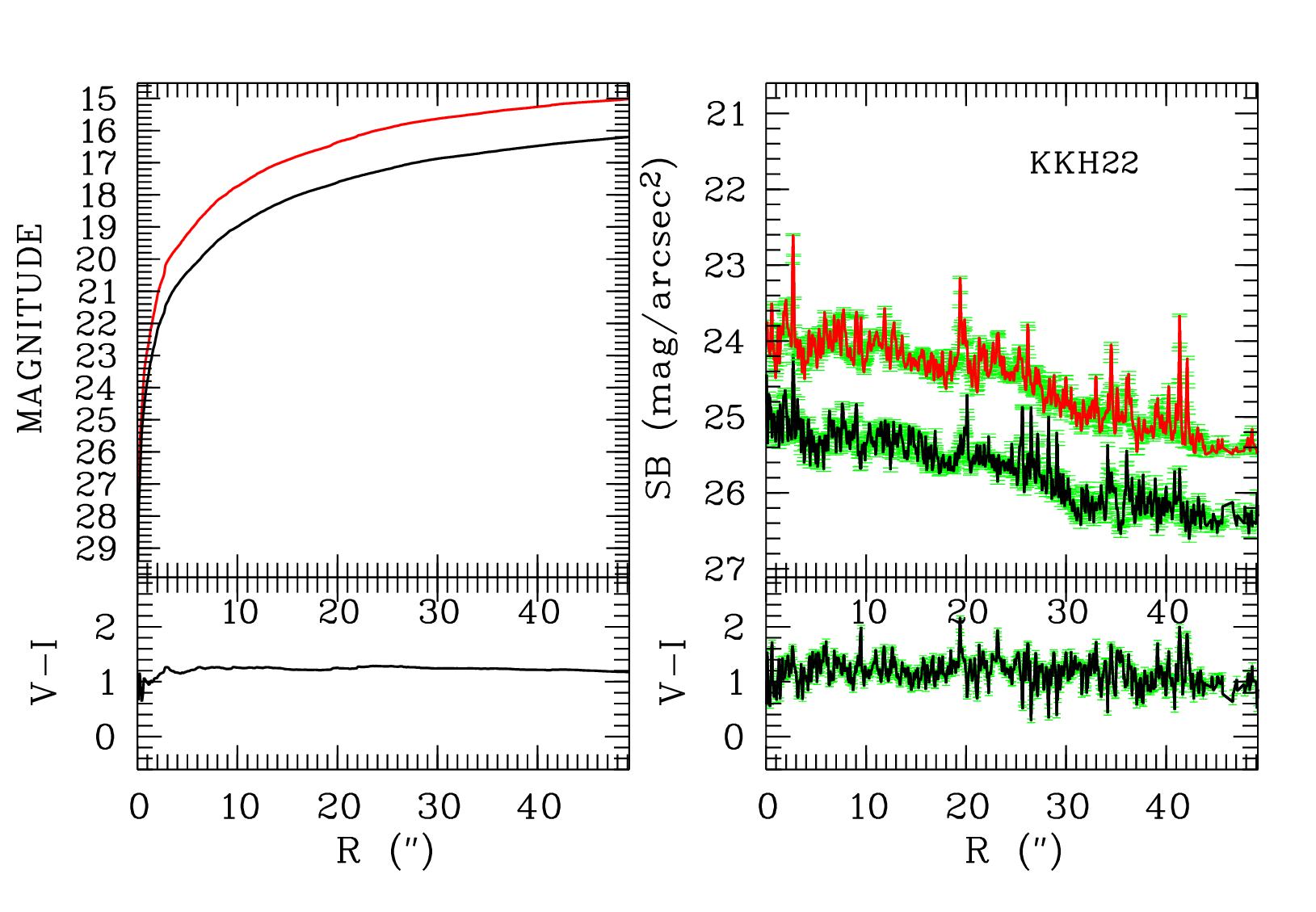}
\end{minipage}
\vfill
\begin{minipage}[h]{0.47\linewidth}
\includegraphics[scale=0.27,angle=-90]{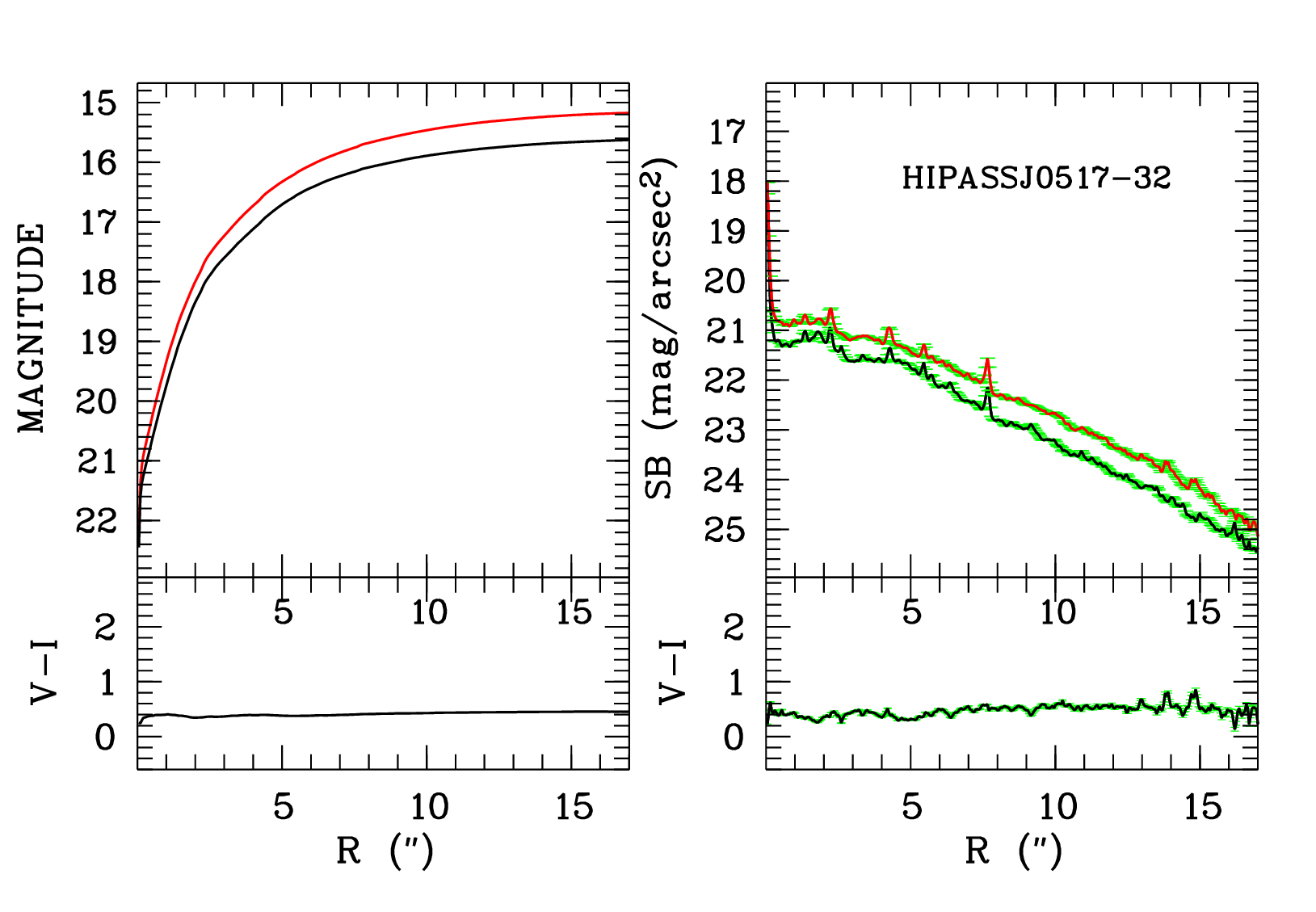}
\end{minipage}
\hfill
\begin{minipage}[h]{0.47\linewidth}
\includegraphics[scale=0.27,angle=-90]{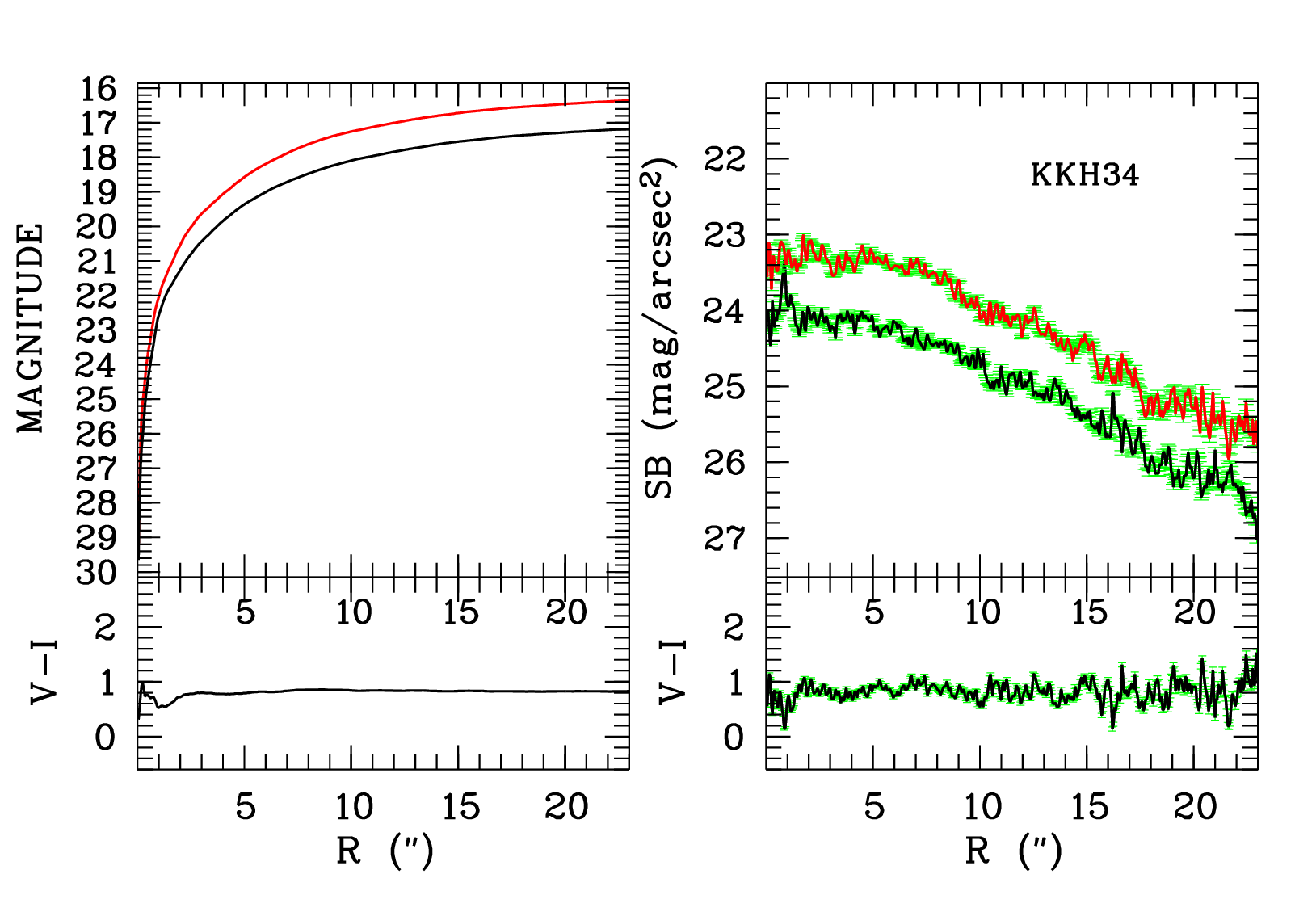}
\end{minipage}
\vfill
\begin{minipage}[h]{0.47\linewidth}
\includegraphics[scale=0.27,angle=-90]{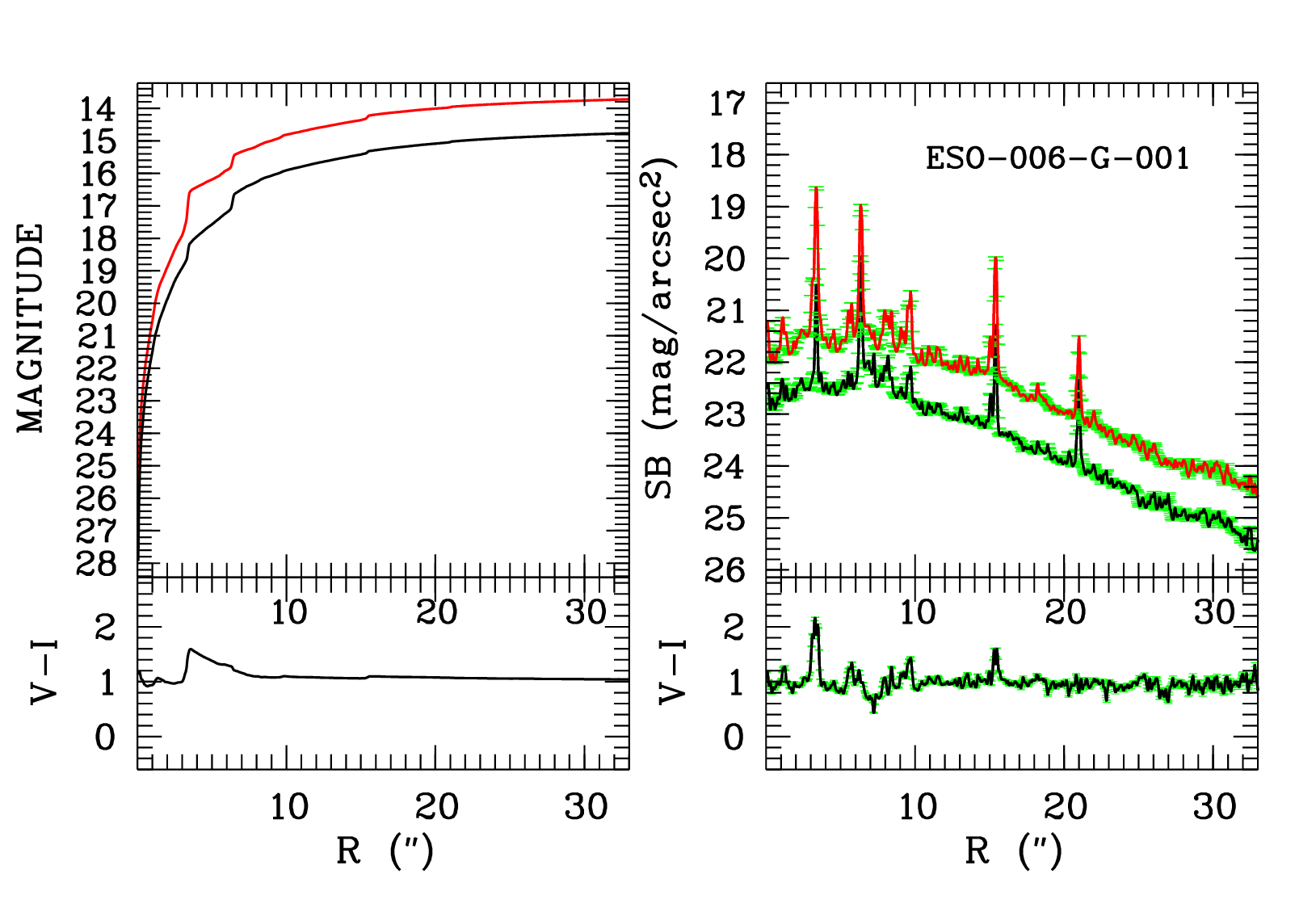}
\end{minipage}
\hfill
\begin{minipage}[h]{0.47\linewidth}
\includegraphics[scale=0.27,angle=-90]{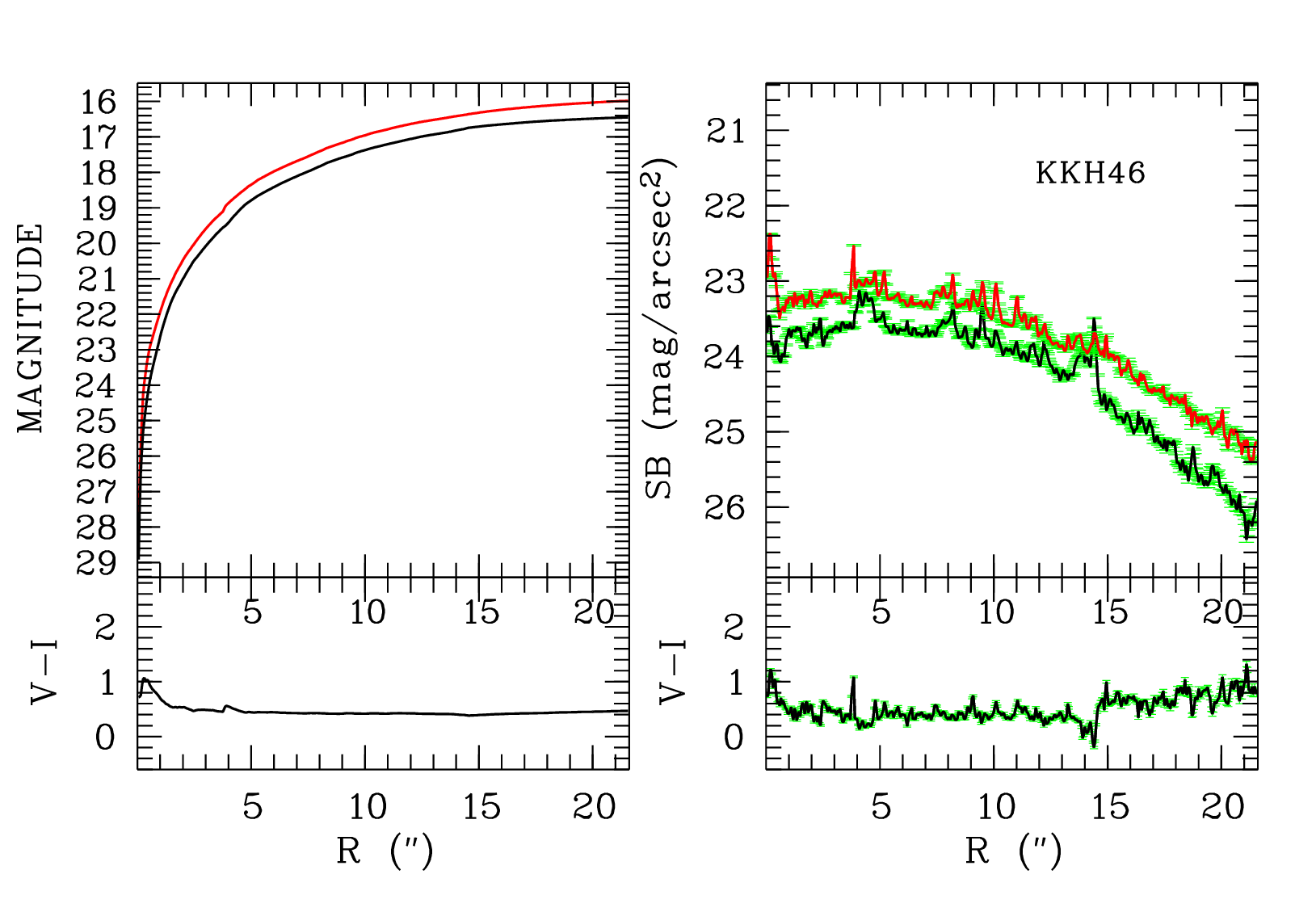}
\end{minipage}
\caption{Continued.}
\end{figure*}
%2
\setcounter{figure}{0}
\begin{figure*}
\begin{minipage}[h]{0.47\linewidth}
\includegraphics[scale=0.27,angle=-90]{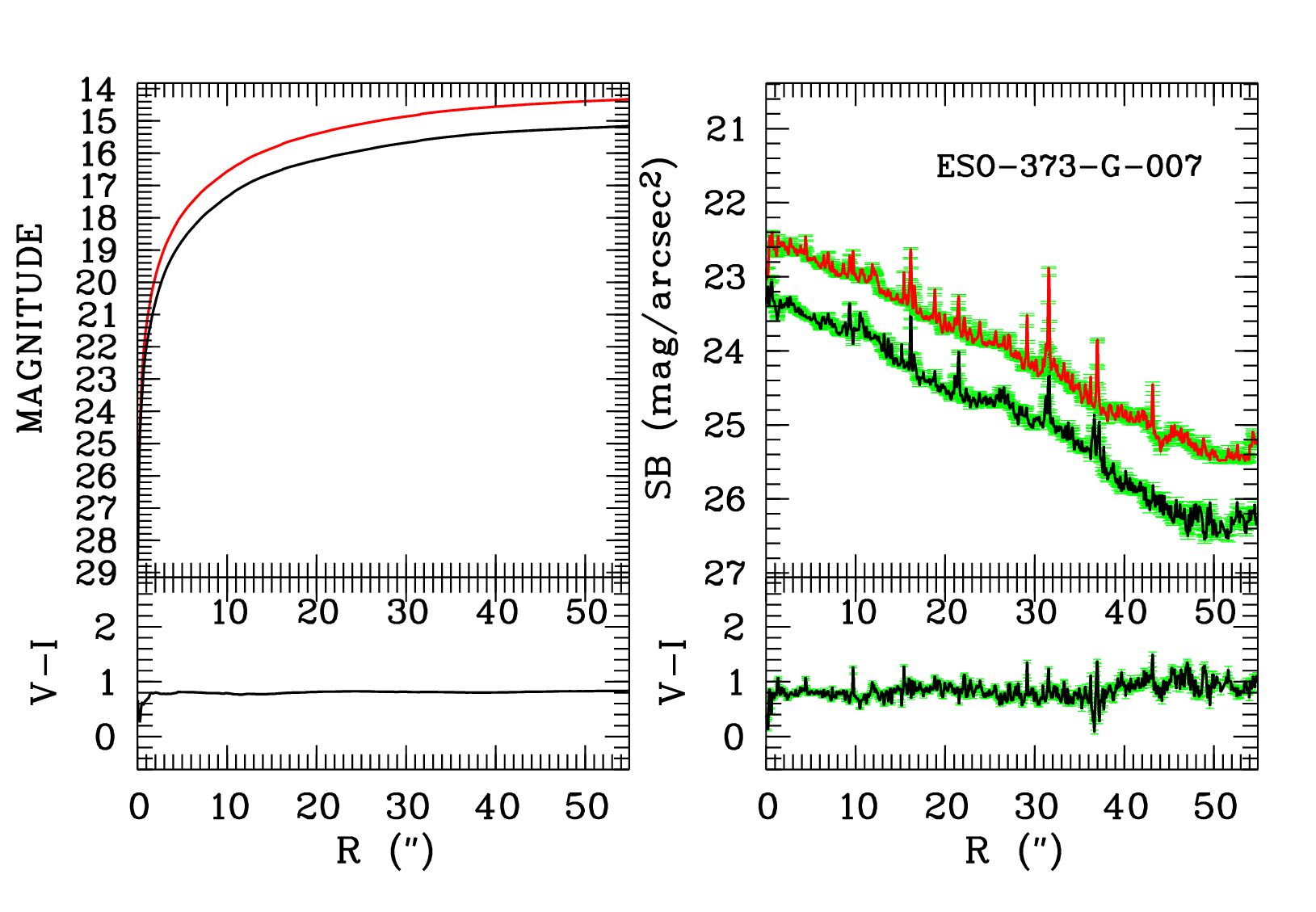}
\end{minipage}
\hfill
\begin{minipage}[h]{0.47\linewidth}
\includegraphics[scale=0.27,angle=-90]{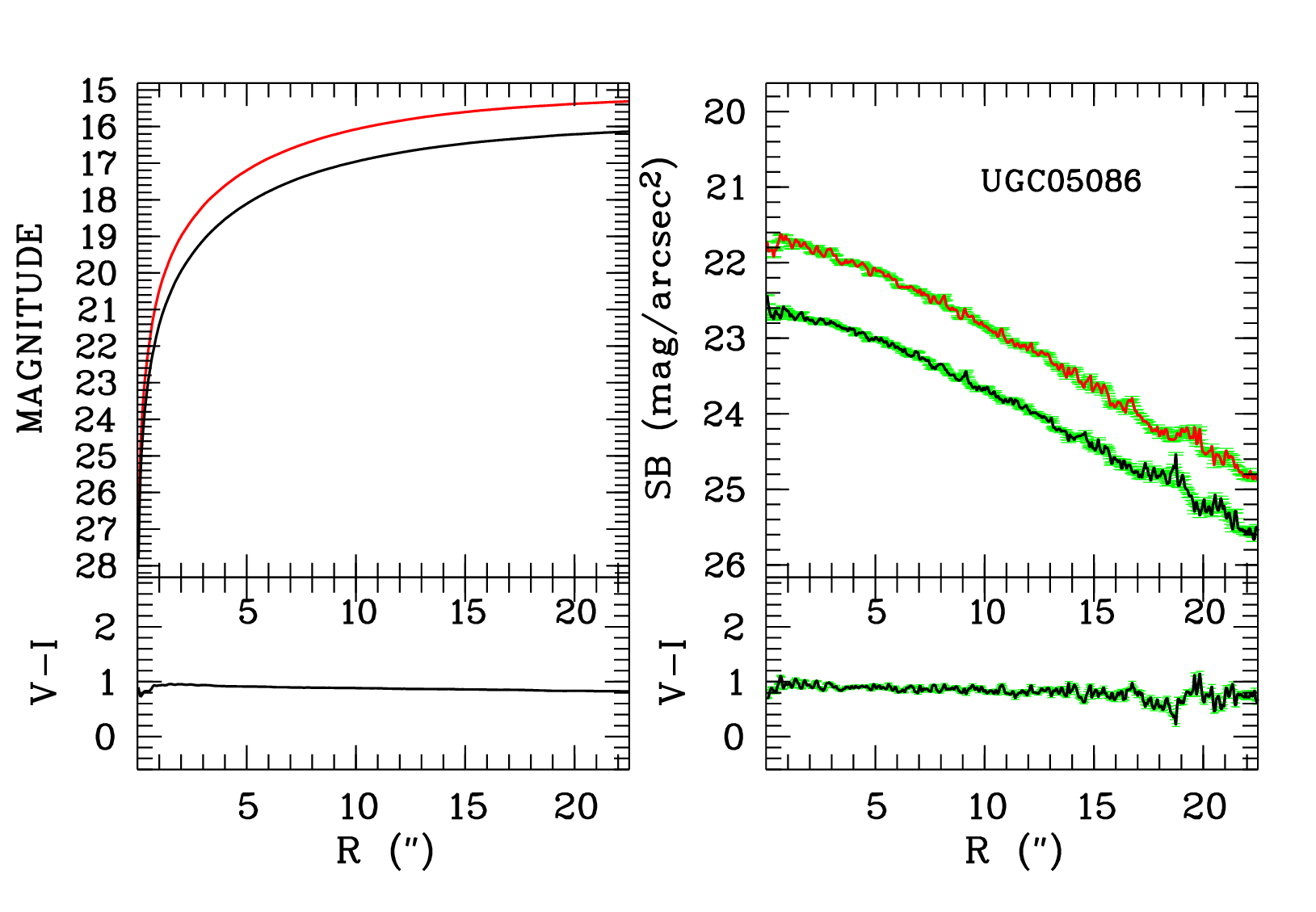}
\end{minipage}
\vfill
\begin{minipage}[h]{0.47\linewidth}
\includegraphics[scale=0.27,angle=-90]{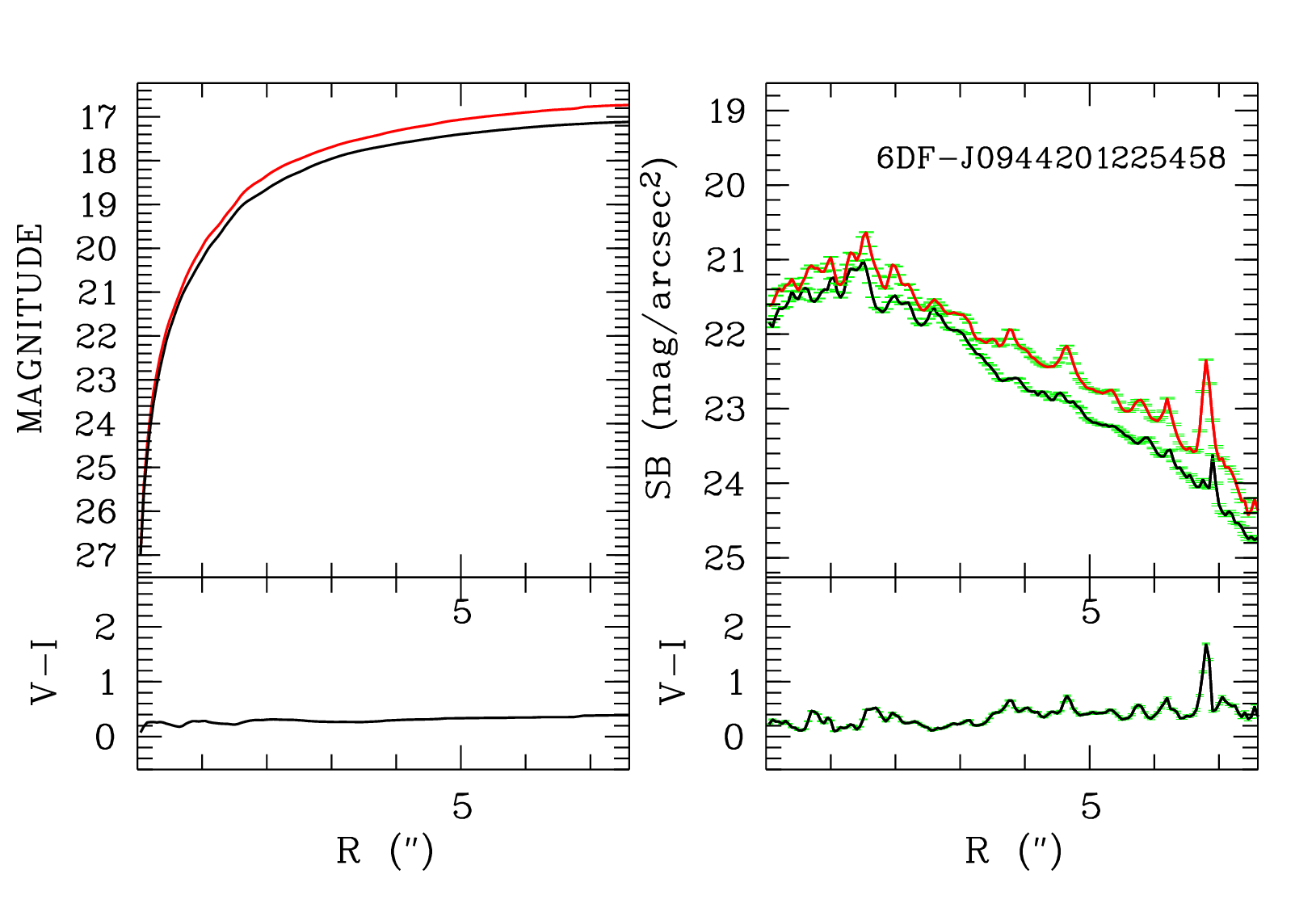}
\end{minipage}
\hfill
\begin{minipage}[h]{0.47\linewidth}
\includegraphics[scale=0.27,angle=-90]{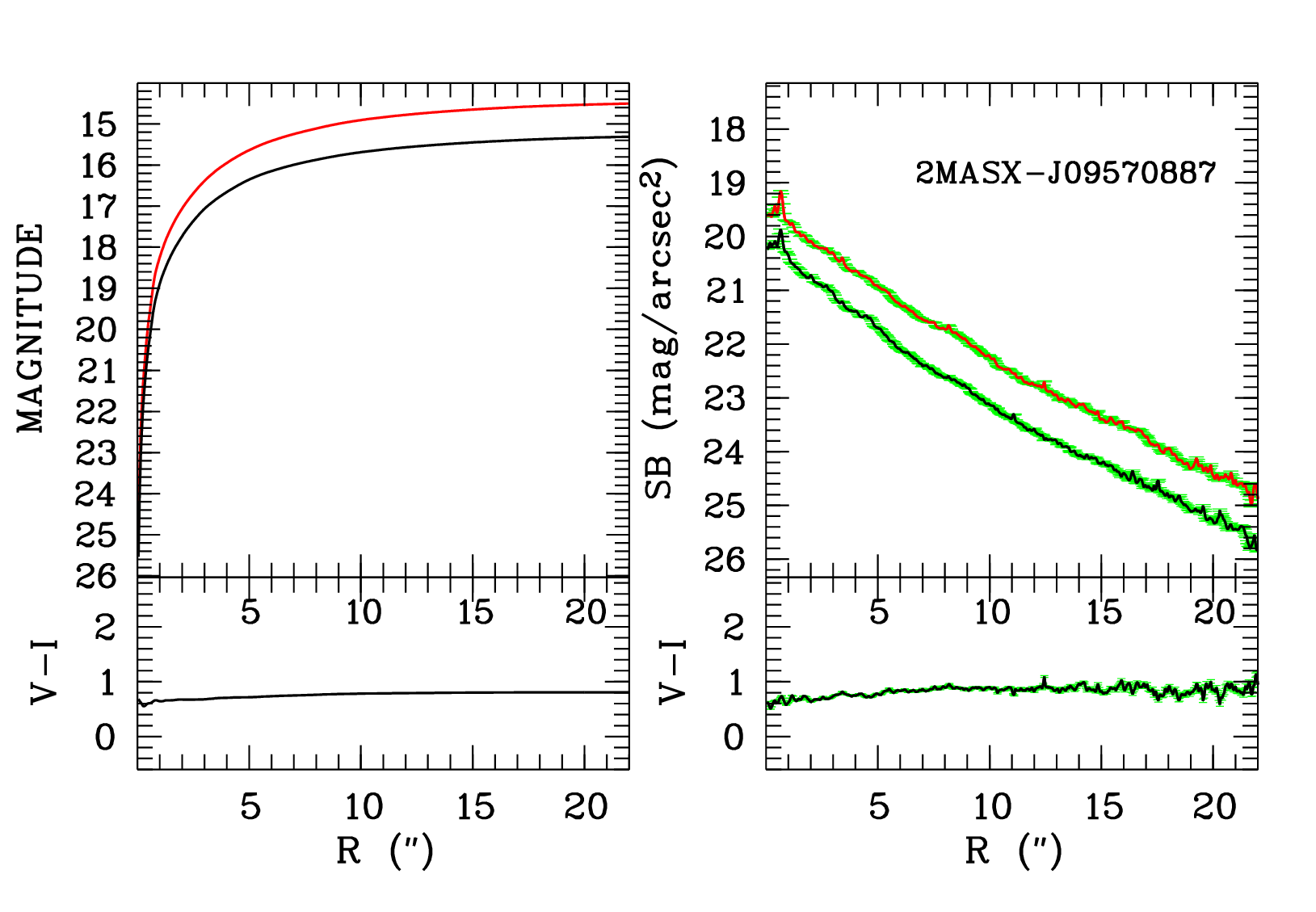}
\end{minipage}
\vfill
\begin{minipage}[h]{0.47\linewidth}
\includegraphics[scale=0.27,angle=-90]{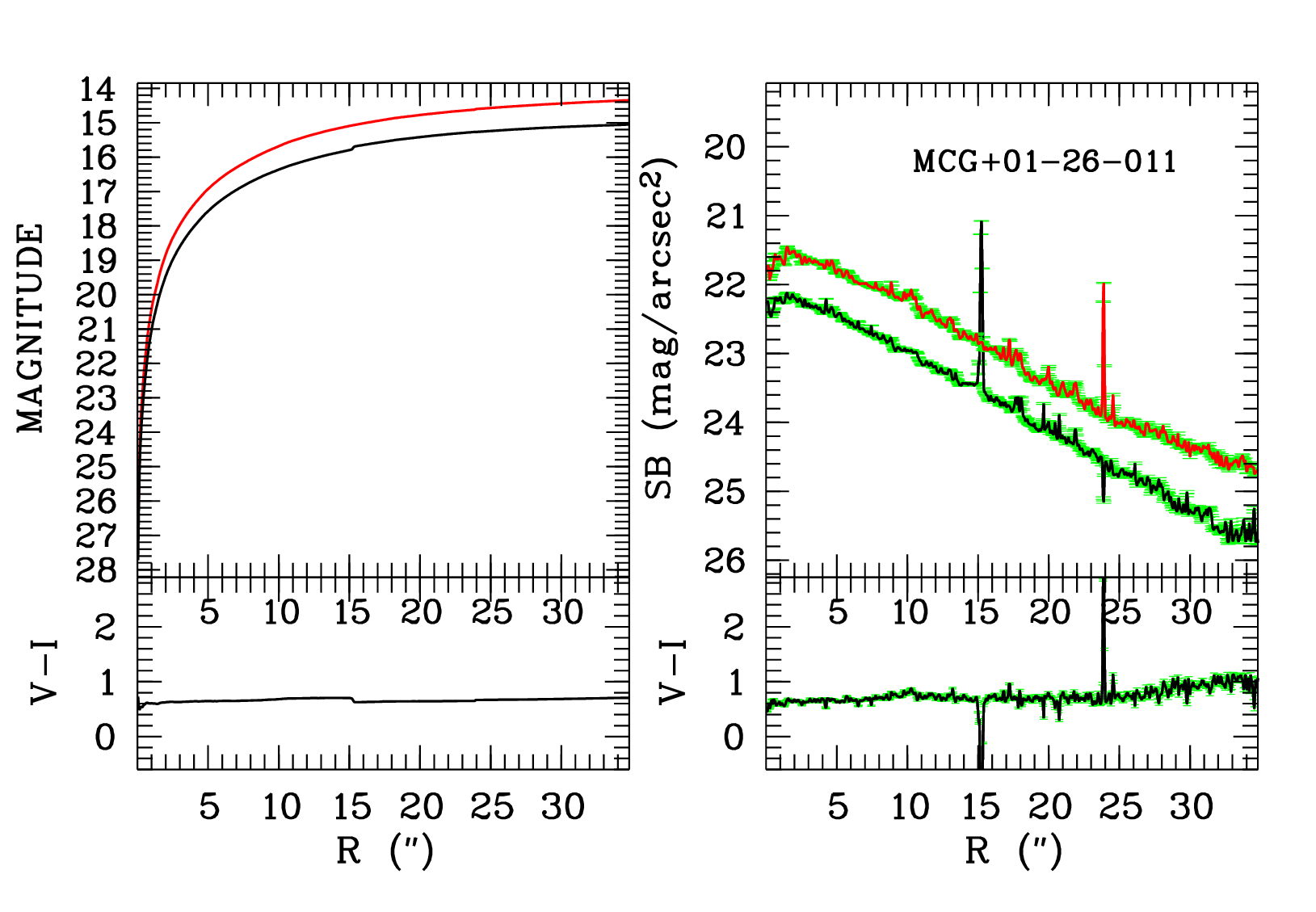}
\end{minipage}
\hfill
\begin{minipage}[h]{0.47\linewidth}
\includegraphics[scale=0.27,angle=-90]{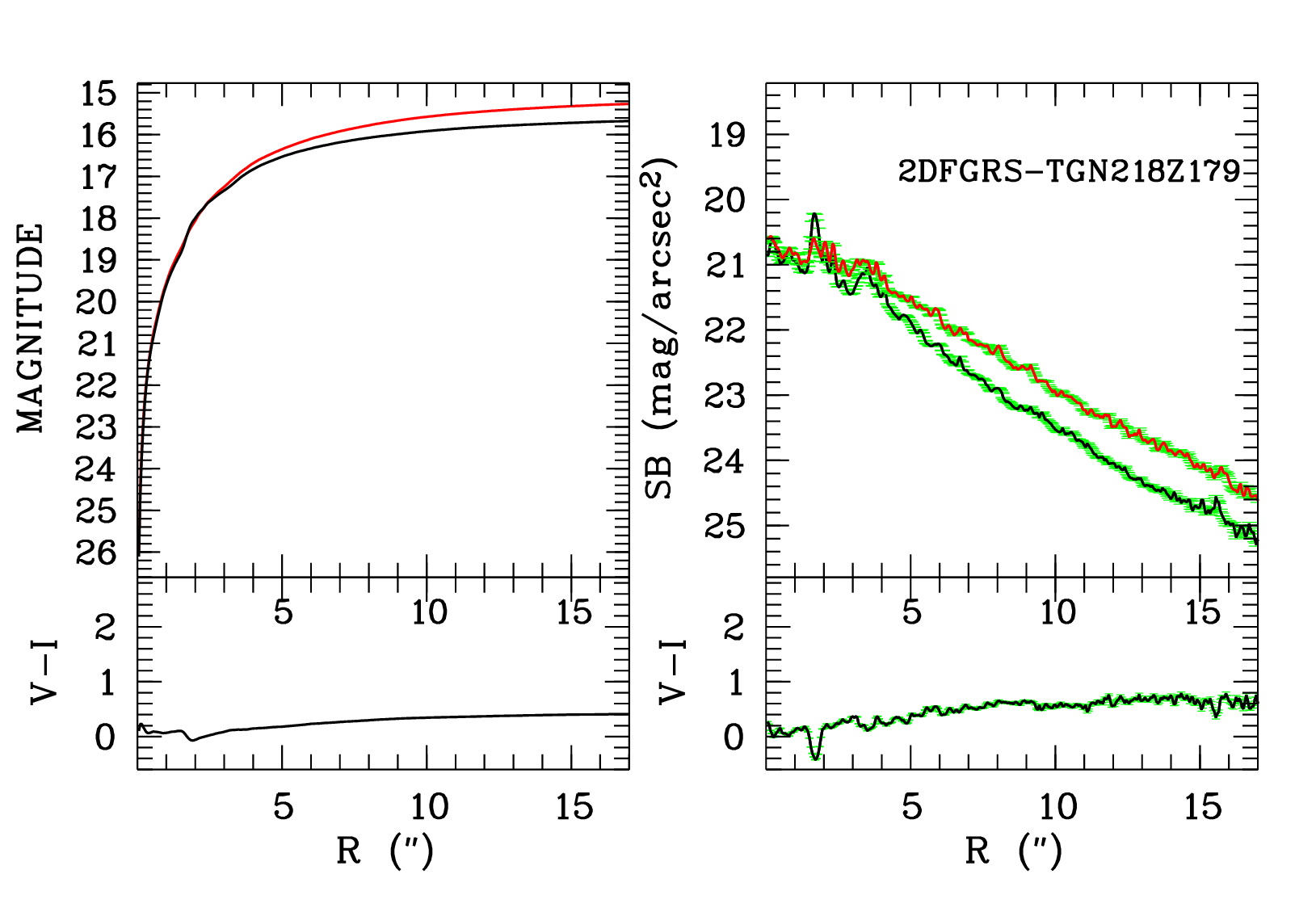}
\end{minipage}
\vfill
\begin{minipage}[h]{0.47\linewidth}
\includegraphics[scale=0.27,angle=-90]{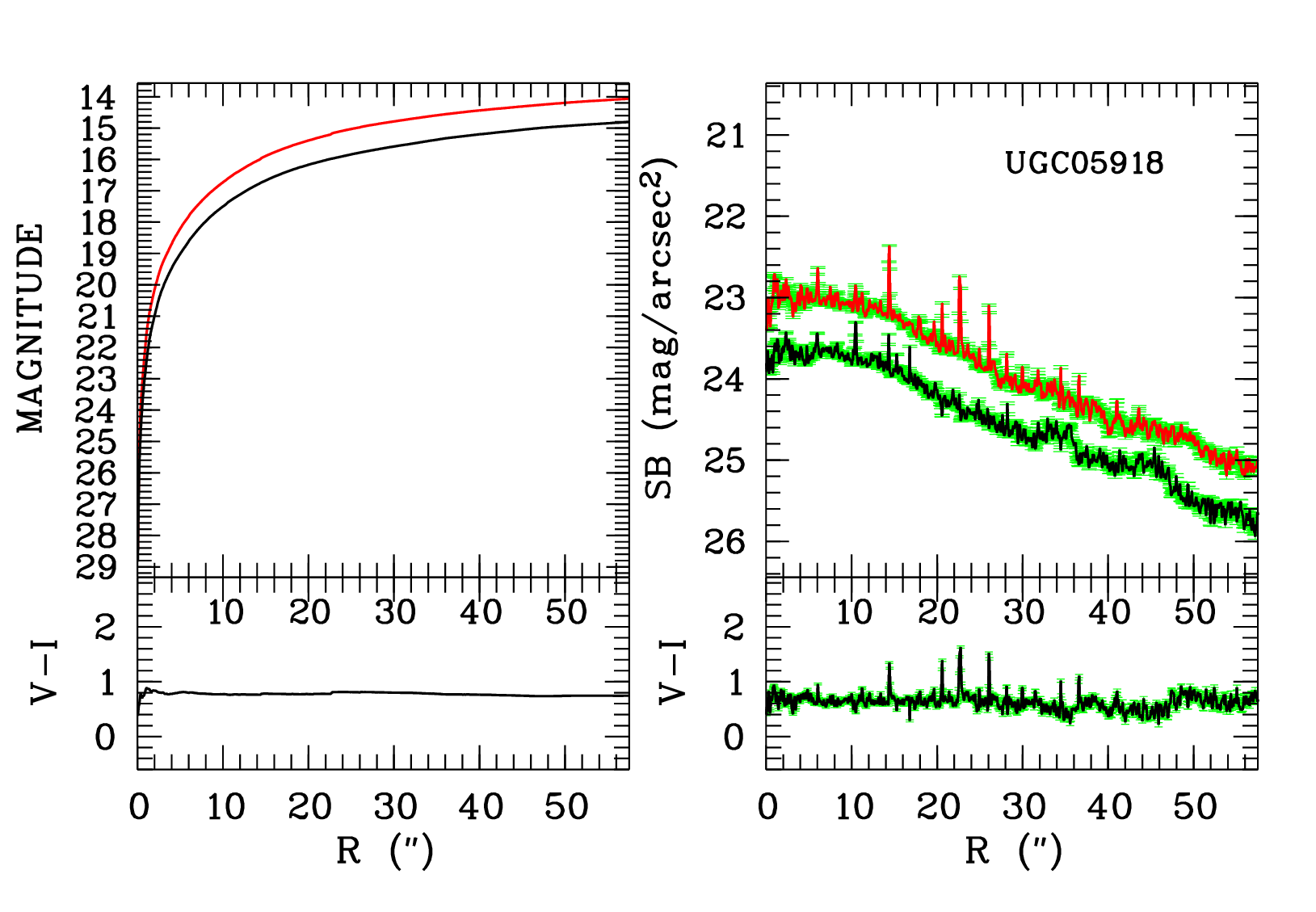}
\end{minipage}
\hfill
\begin{minipage}[h]{0.47\linewidth}
\includegraphics[scale=0.27,angle=-90]{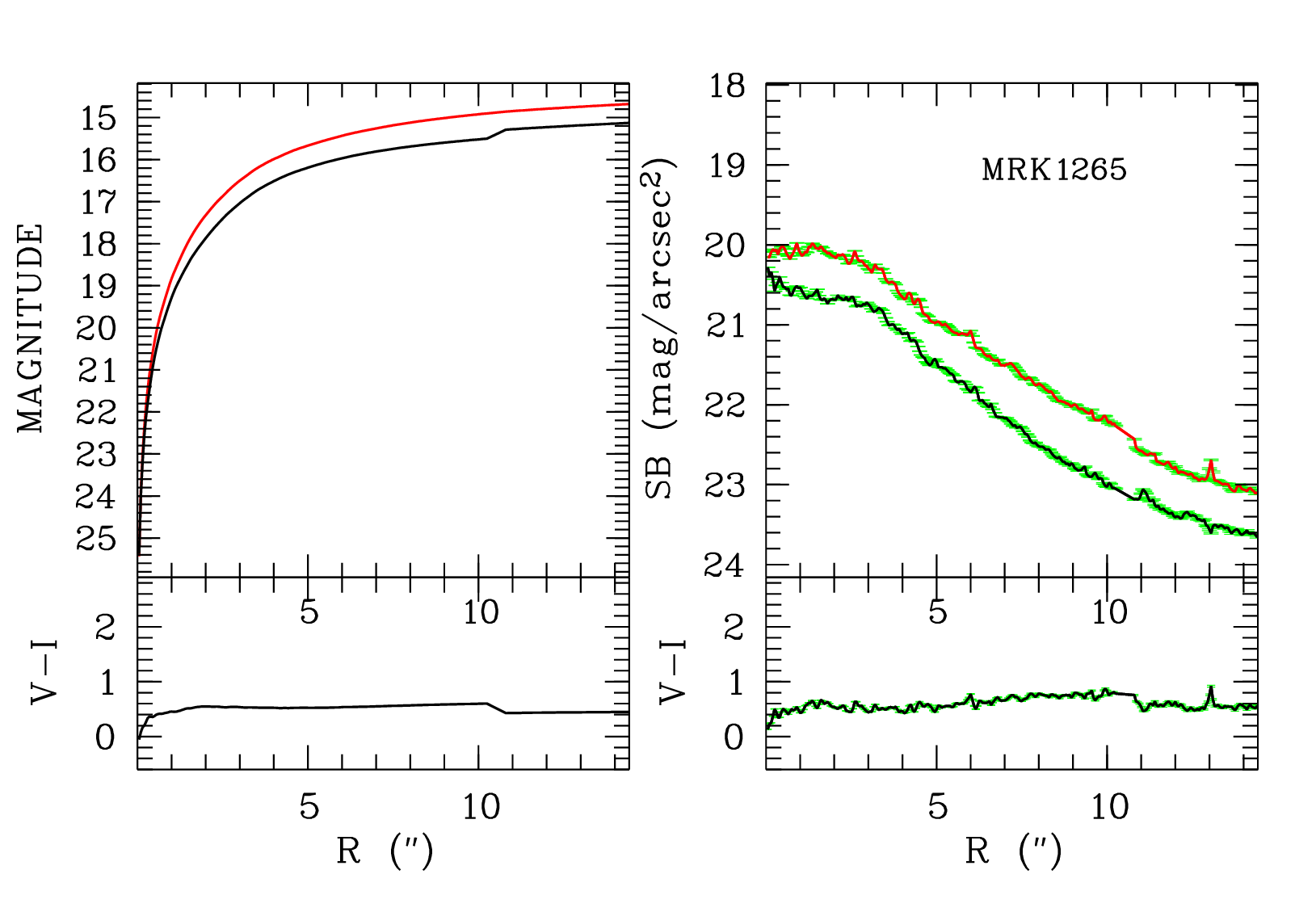}
\end{minipage}
\caption{Continued.}
\end{figure*}

\setcounter{figure}{0}
\begin{figure*}
\begin{minipage}[h]{0.47\linewidth}
\includegraphics[scale=0.27,angle=-90]{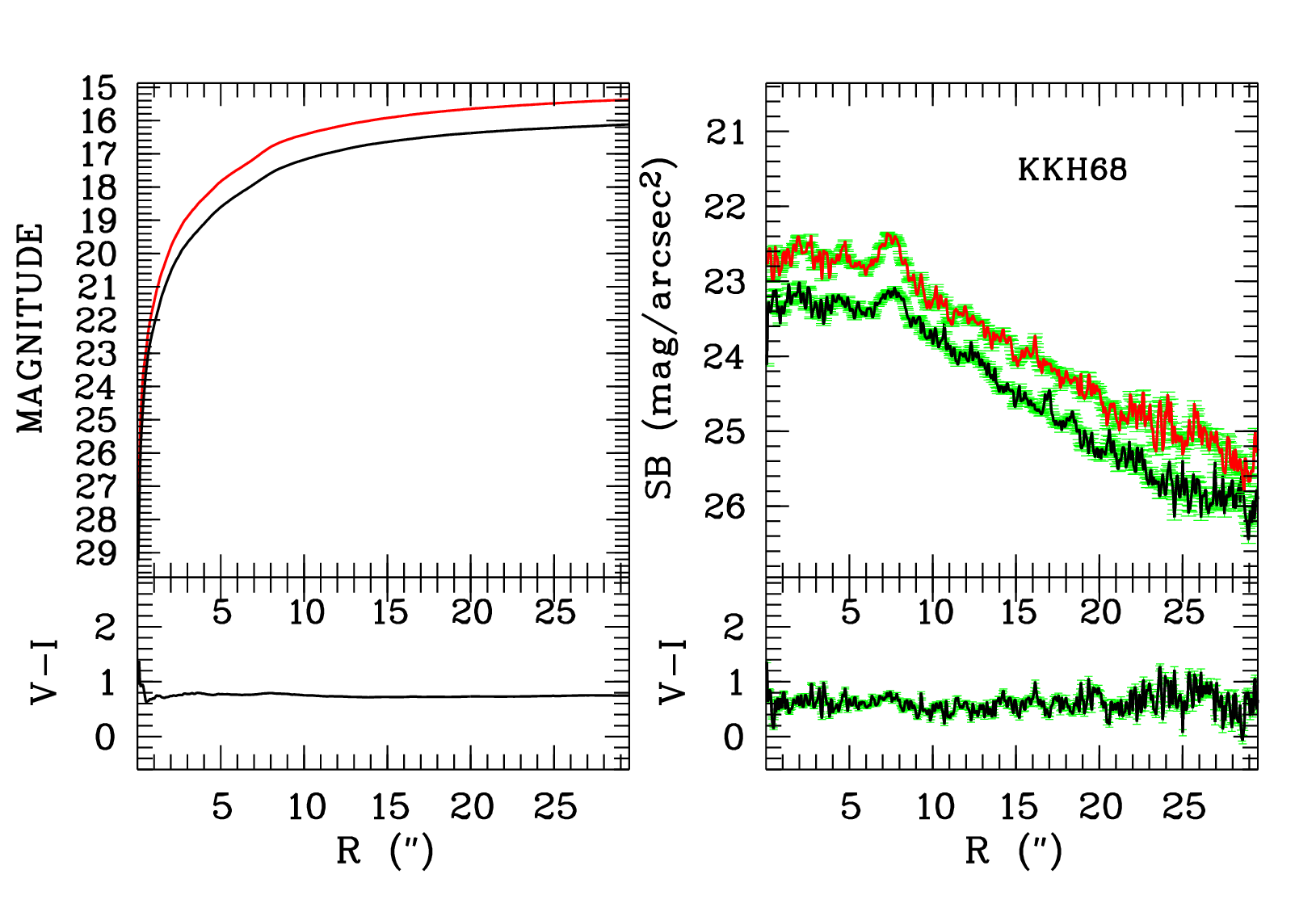}
\end{minipage}
\hfill
\begin{minipage}[h]{0.47\linewidth}
\includegraphics[scale=0.27,angle=-90]{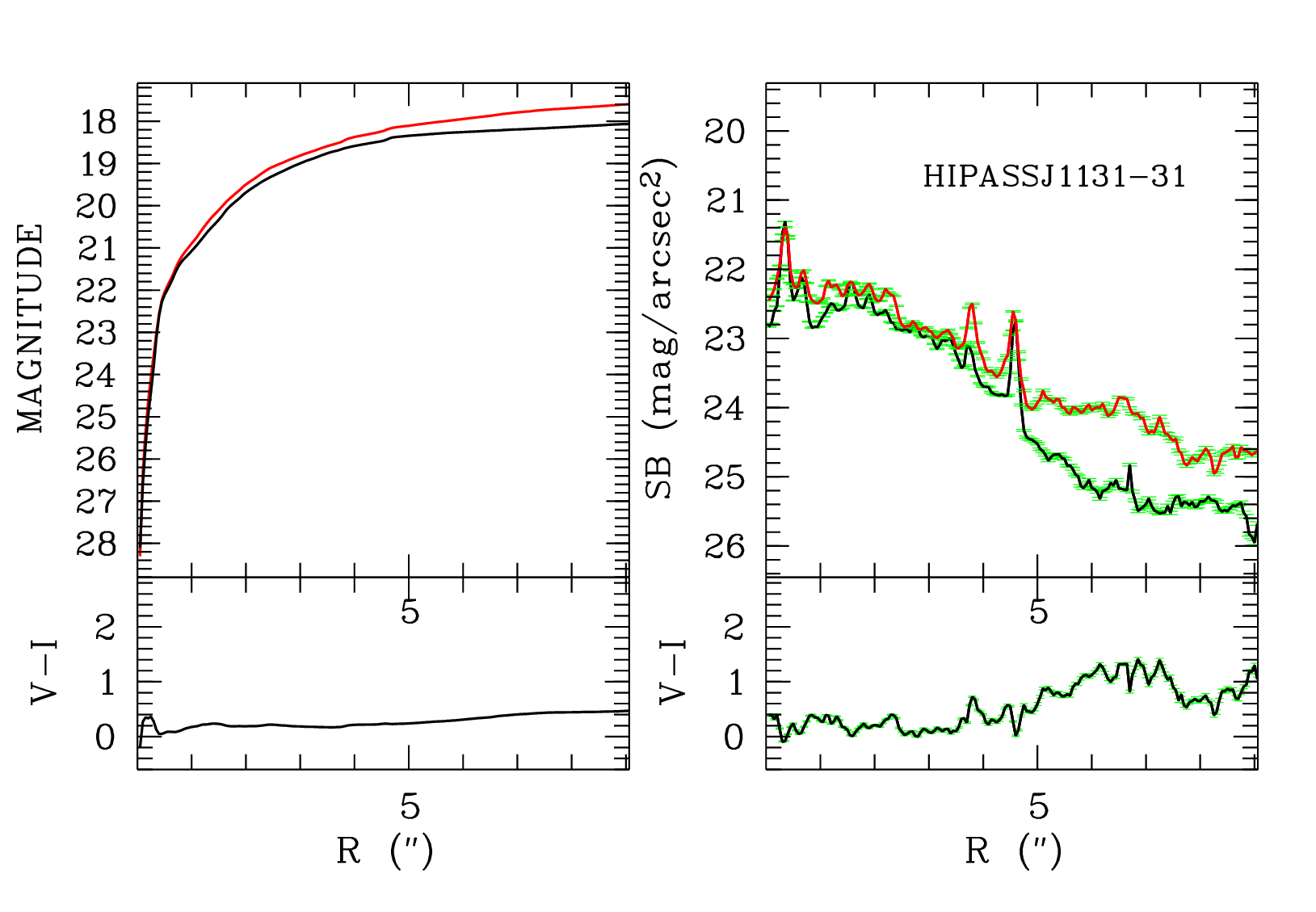}
\end{minipage}
\vfill
\begin{minipage}[h]{0.47\linewidth}
\includegraphics[scale=0.27,angle=-90]{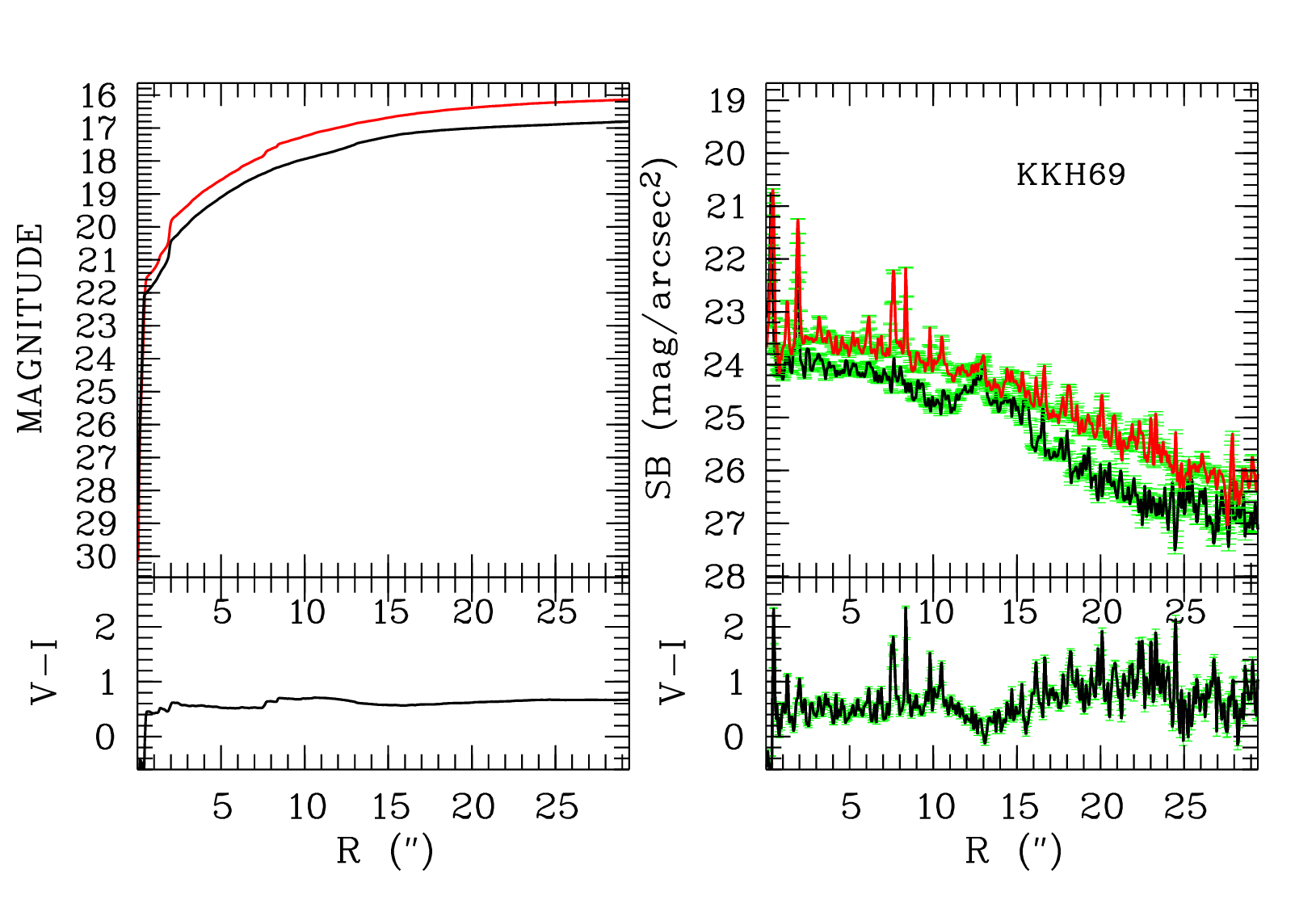}
\end{minipage}
\hfill
\begin{minipage}[h]{0.47\linewidth}
\includegraphics[scale=0.27,angle=-90]{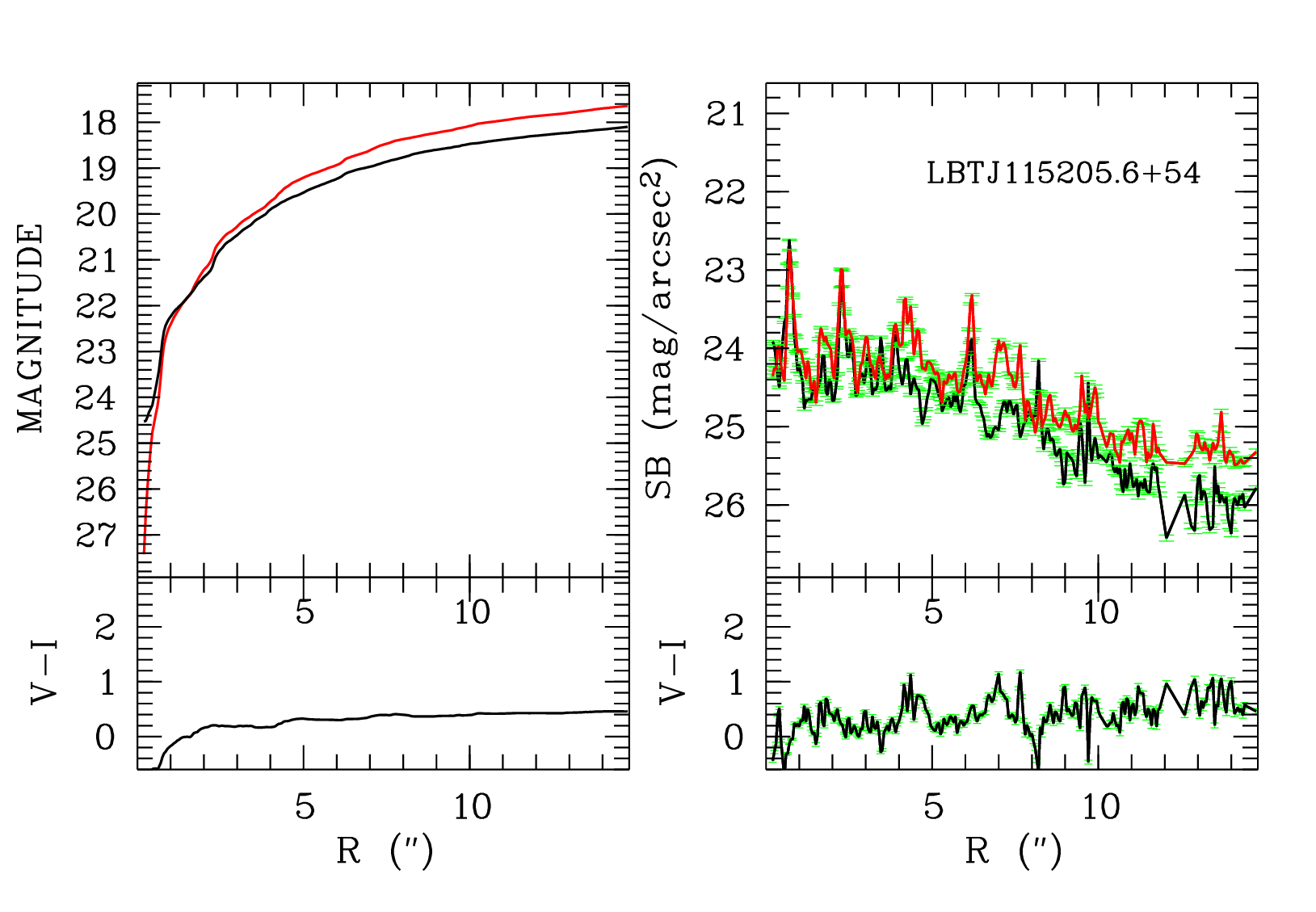}
\end{minipage}
\vfill
\begin{minipage}[h]{0.47\linewidth}
\includegraphics[scale=0.27,angle=-90]{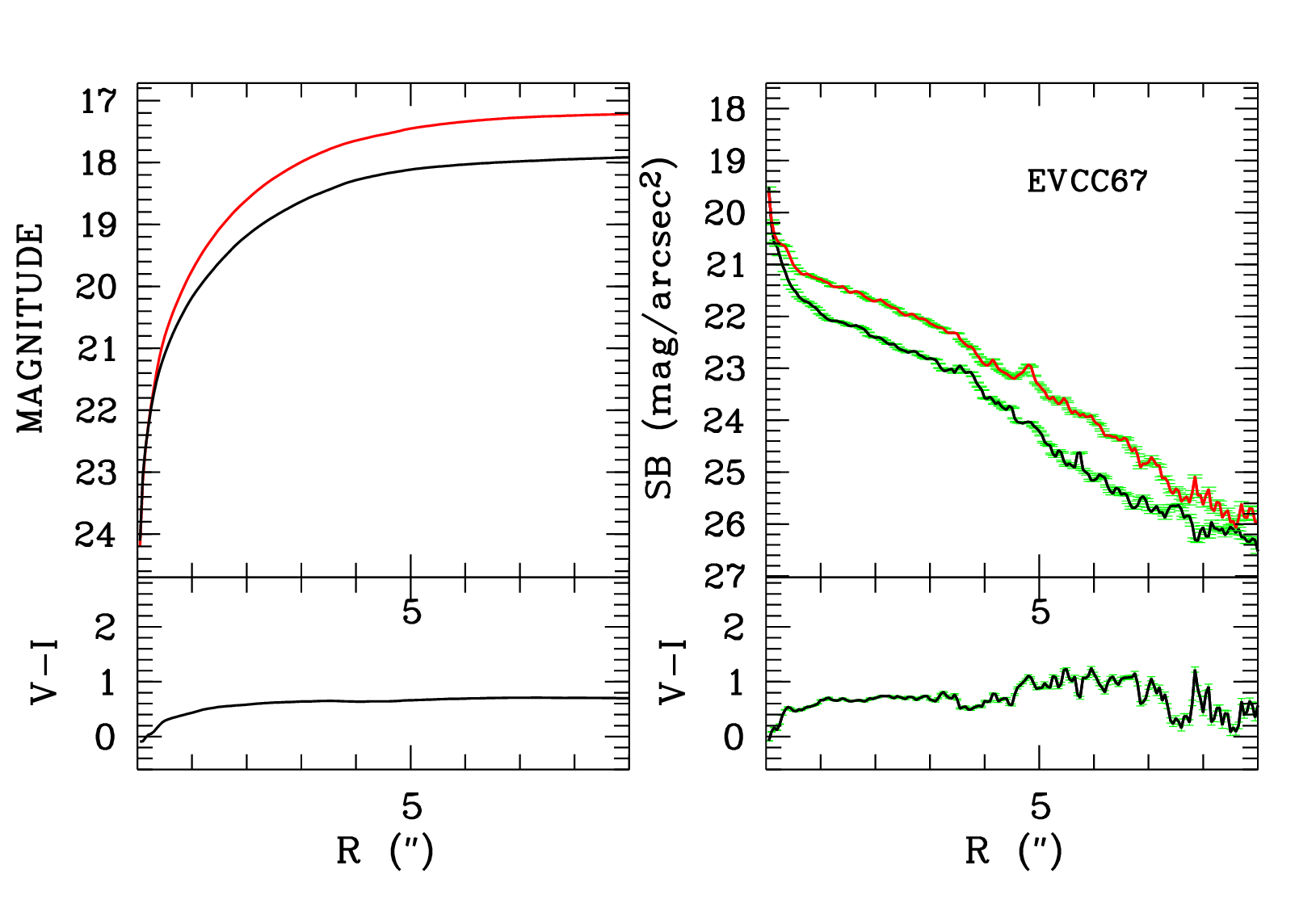}
\end{minipage}
\hfill
\begin{minipage}[h]{0.47\linewidth}
\includegraphics[scale=0.27,angle=-90]{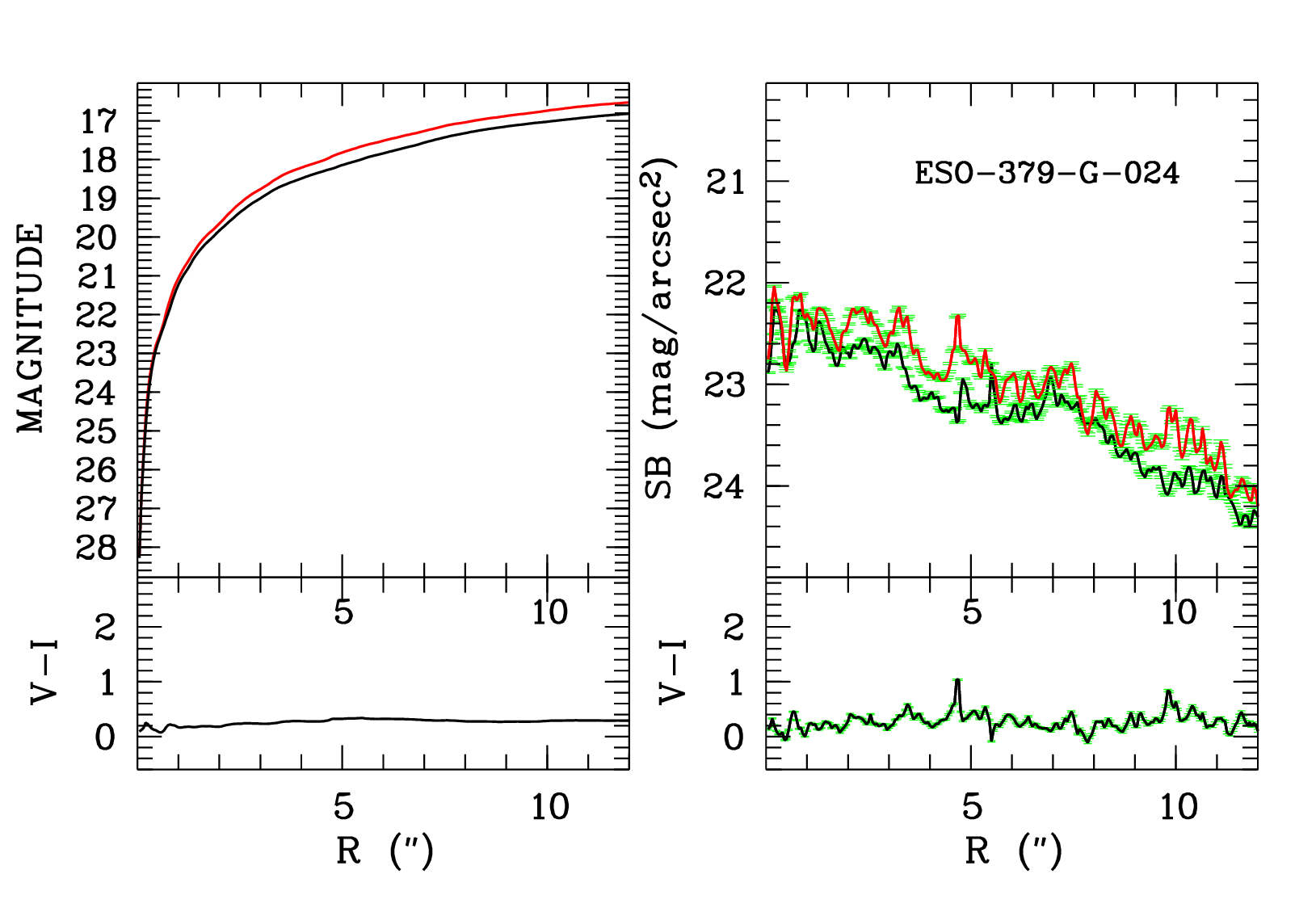}
\end{minipage}
\vfill
\begin{minipage}[h]{0.47\linewidth}
\includegraphics[scale=0.27,angle=-90]{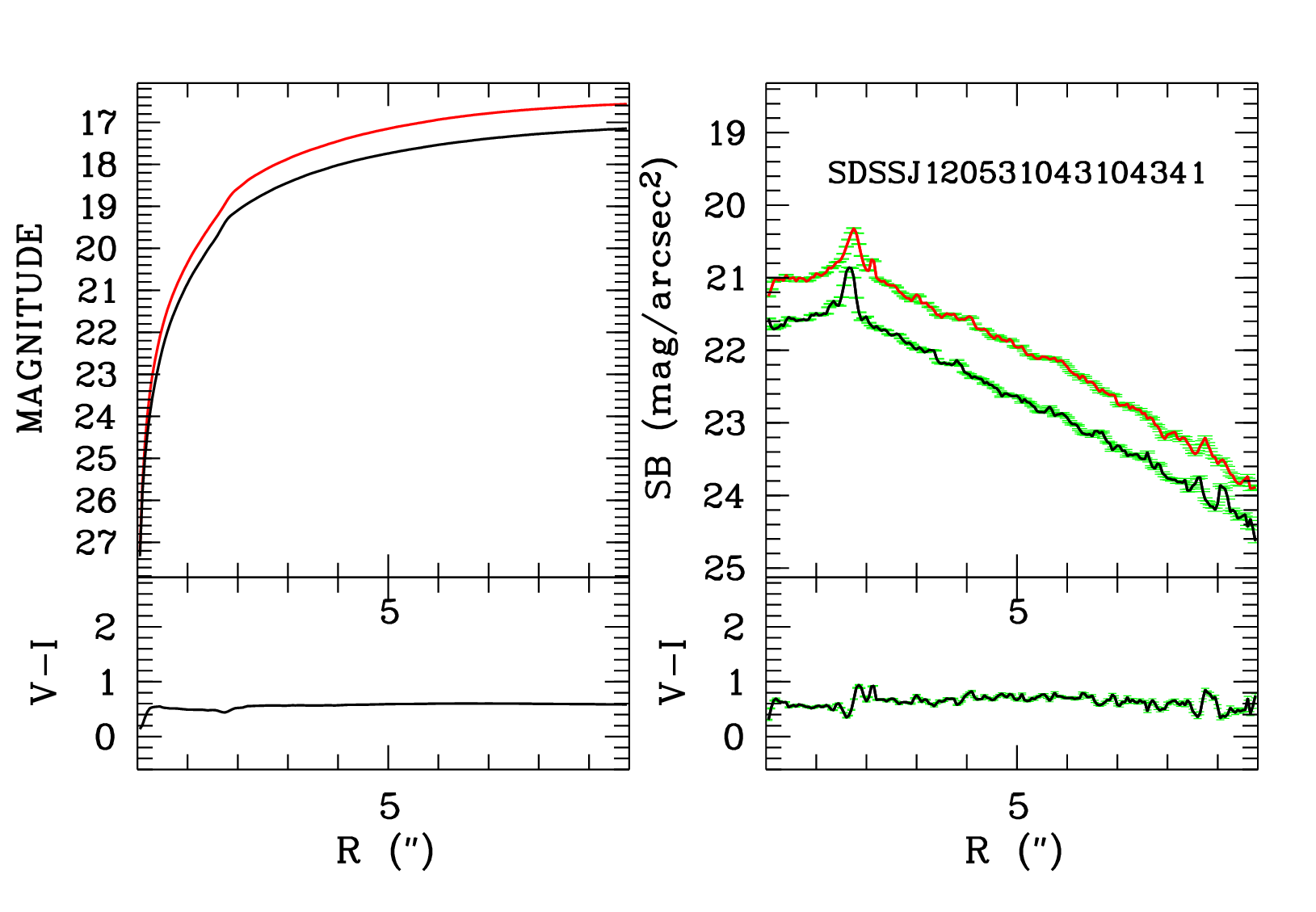}
\end{minipage}
\hfill
\begin{minipage}[h]{0.47\linewidth}
\includegraphics[scale=0.27,angle=-90]{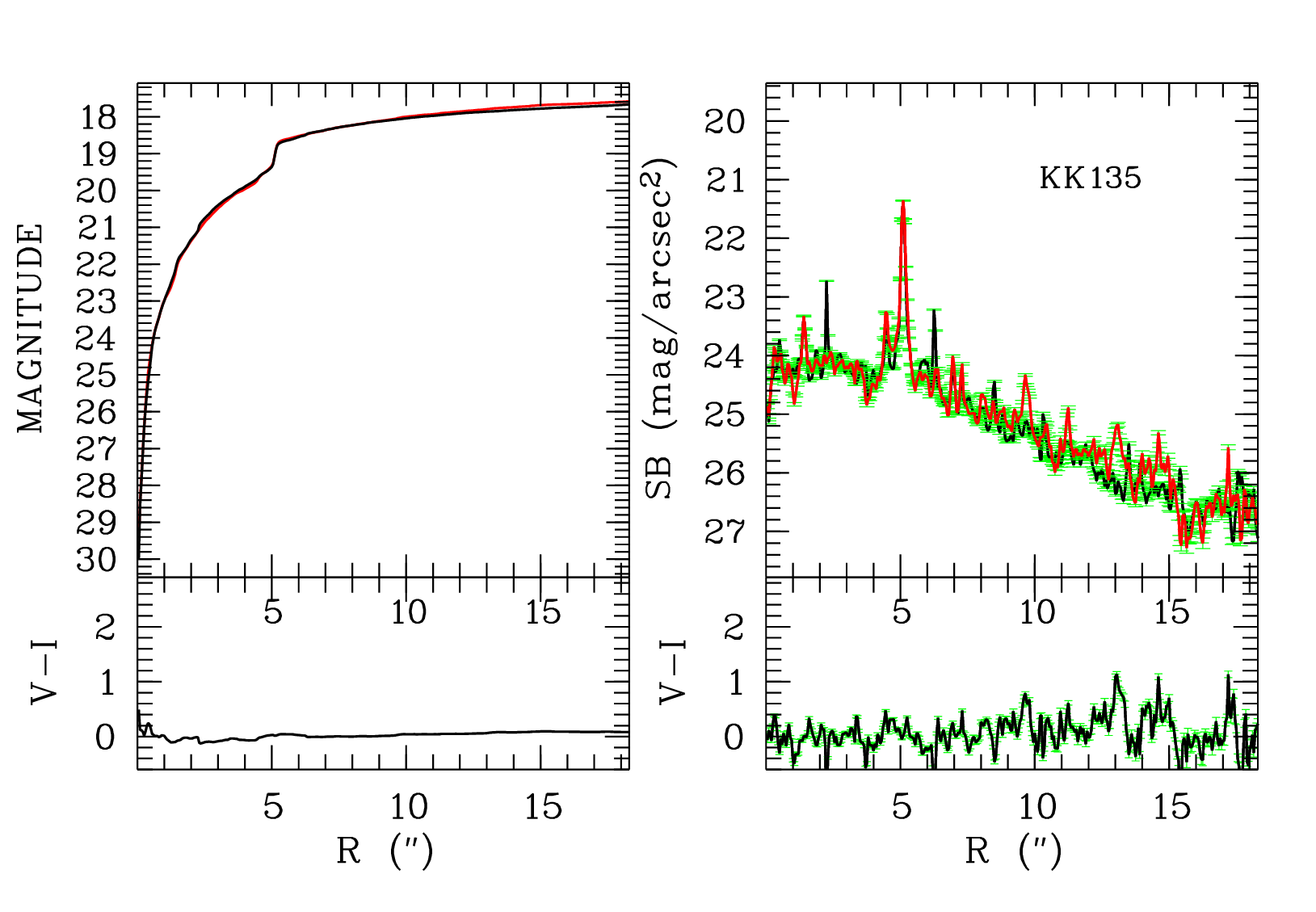}
\end{minipage}
\caption{Continued.}
\end{figure*}

\setcounter{figure}{0}
\begin{figure*}
\begin{minipage}[h]{0.47\linewidth}
\includegraphics[scale=0.27,angle=-90]{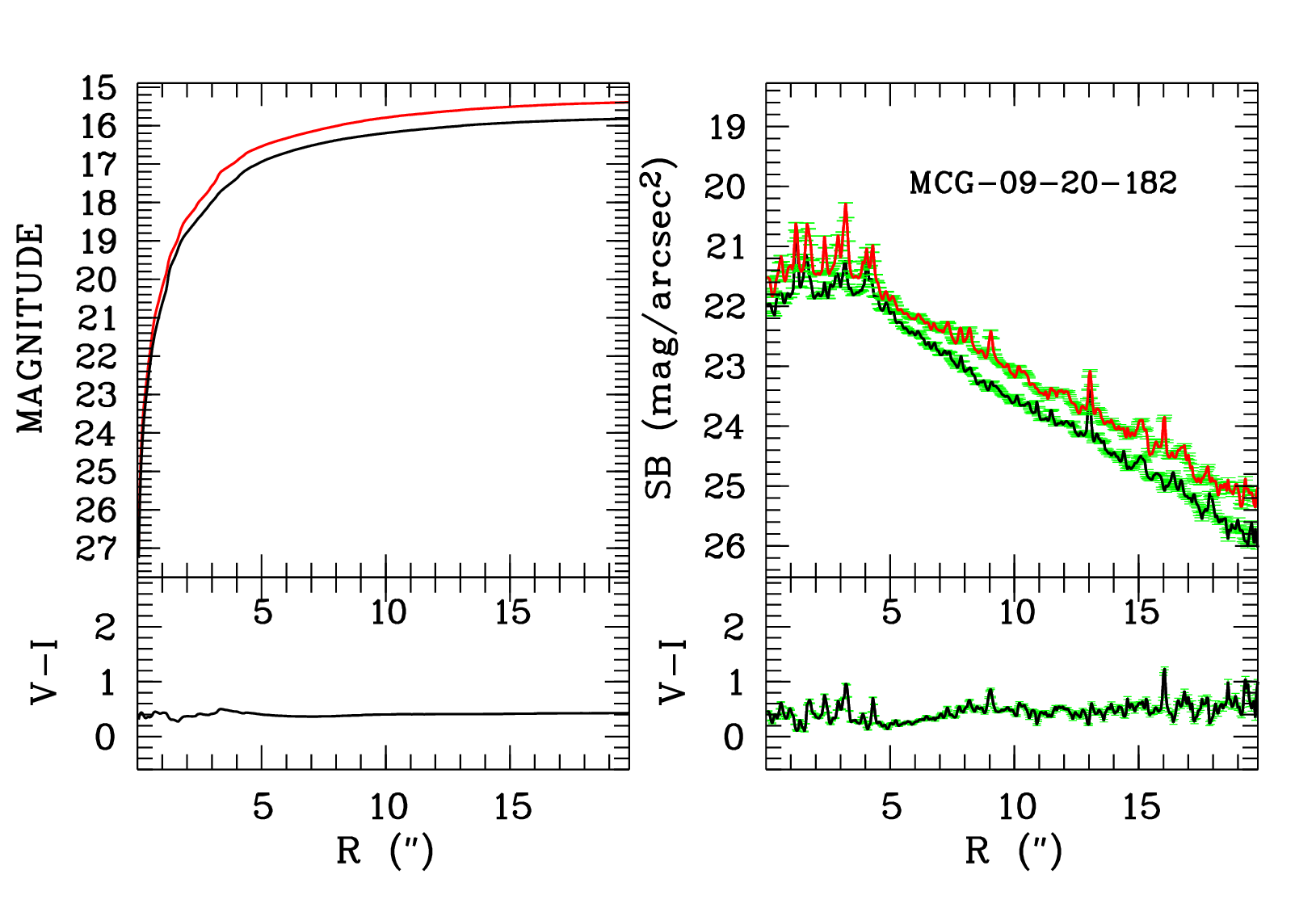}
\end{minipage}
\hfill
\begin{minipage}[h]{0.47\linewidth}
\includegraphics[scale=0.27,angle=-90]{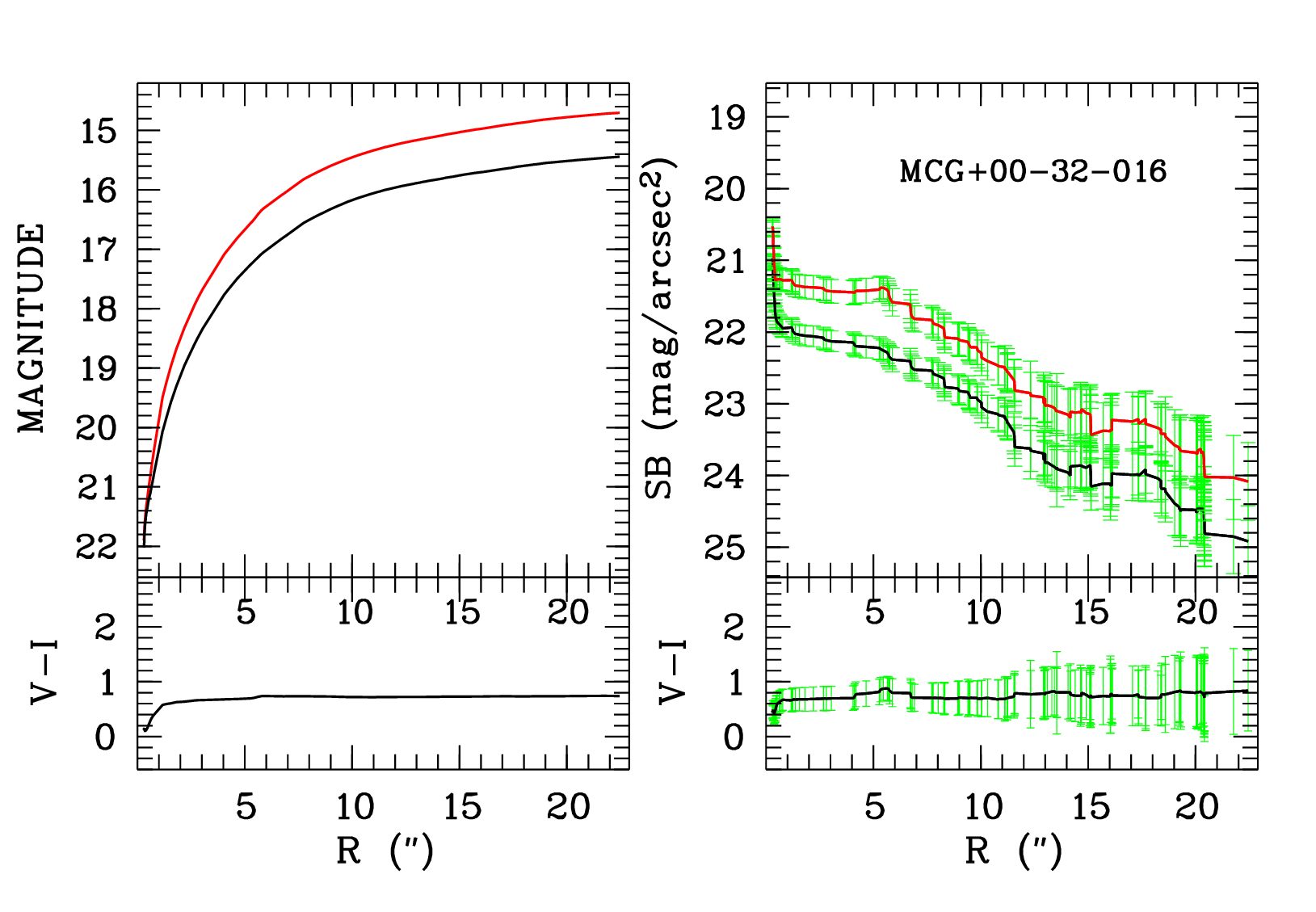}
\end{minipage}
\vfill
\begin{minipage}[h]{0.47\linewidth}
\includegraphics[scale=0.27,angle=-90]{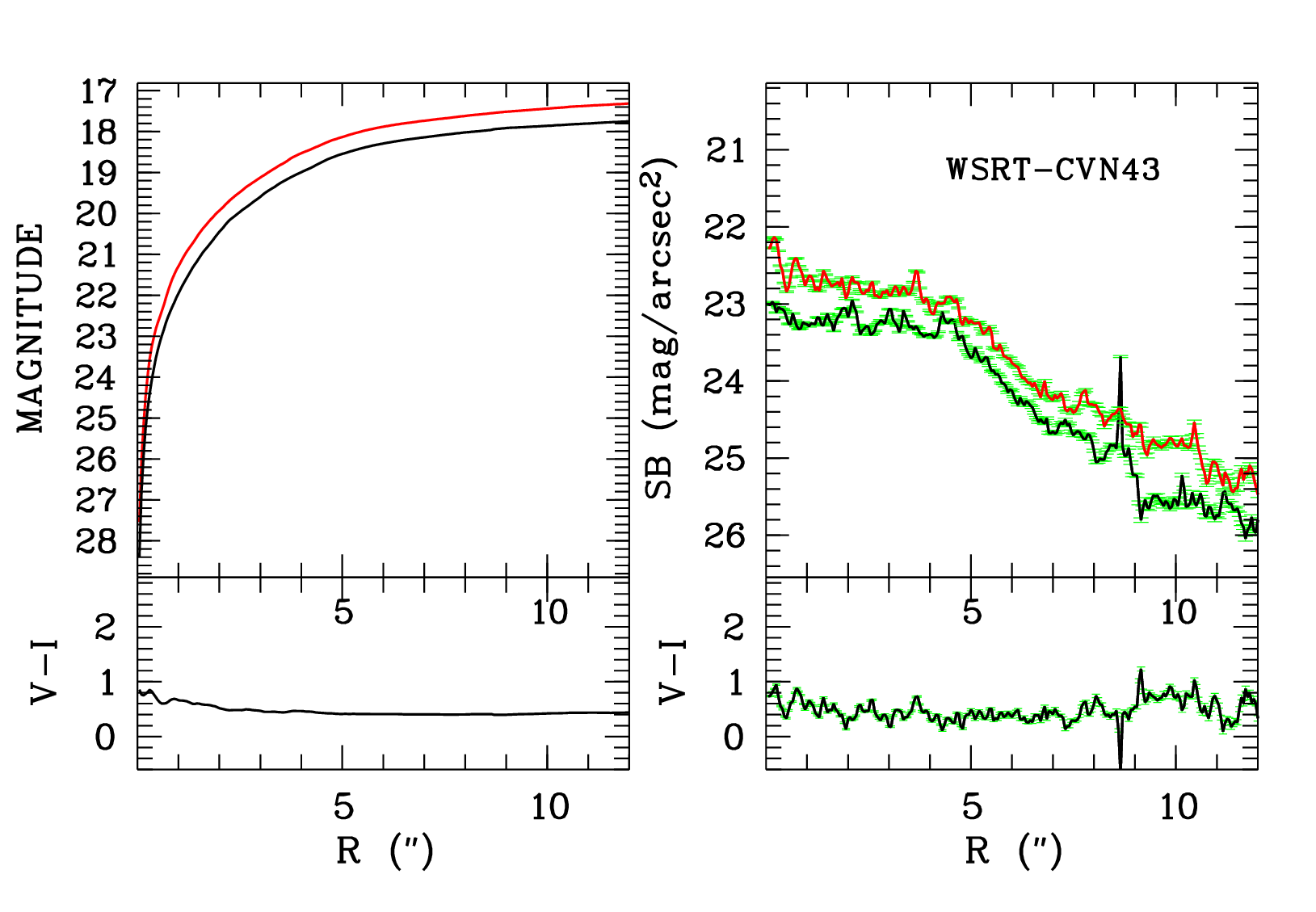}
\end{minipage}
\hfill
\begin{minipage}[h]{0.47\linewidth}
\includegraphics[scale=0.27,angle=-90]{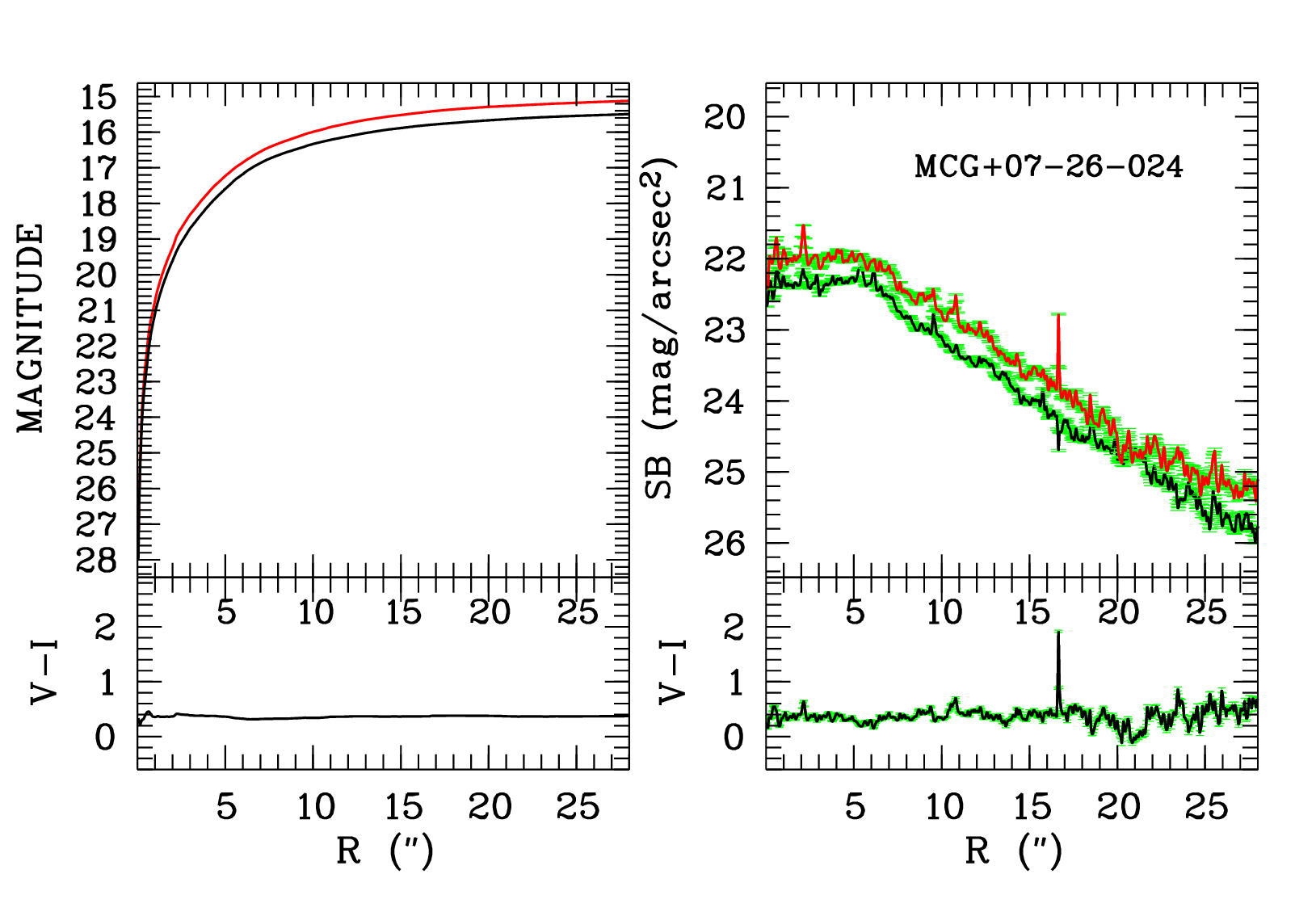}
\end{minipage}
\vfill
\begin{minipage}[h]{0.47\linewidth}
\includegraphics[scale=0.27,angle=-90]{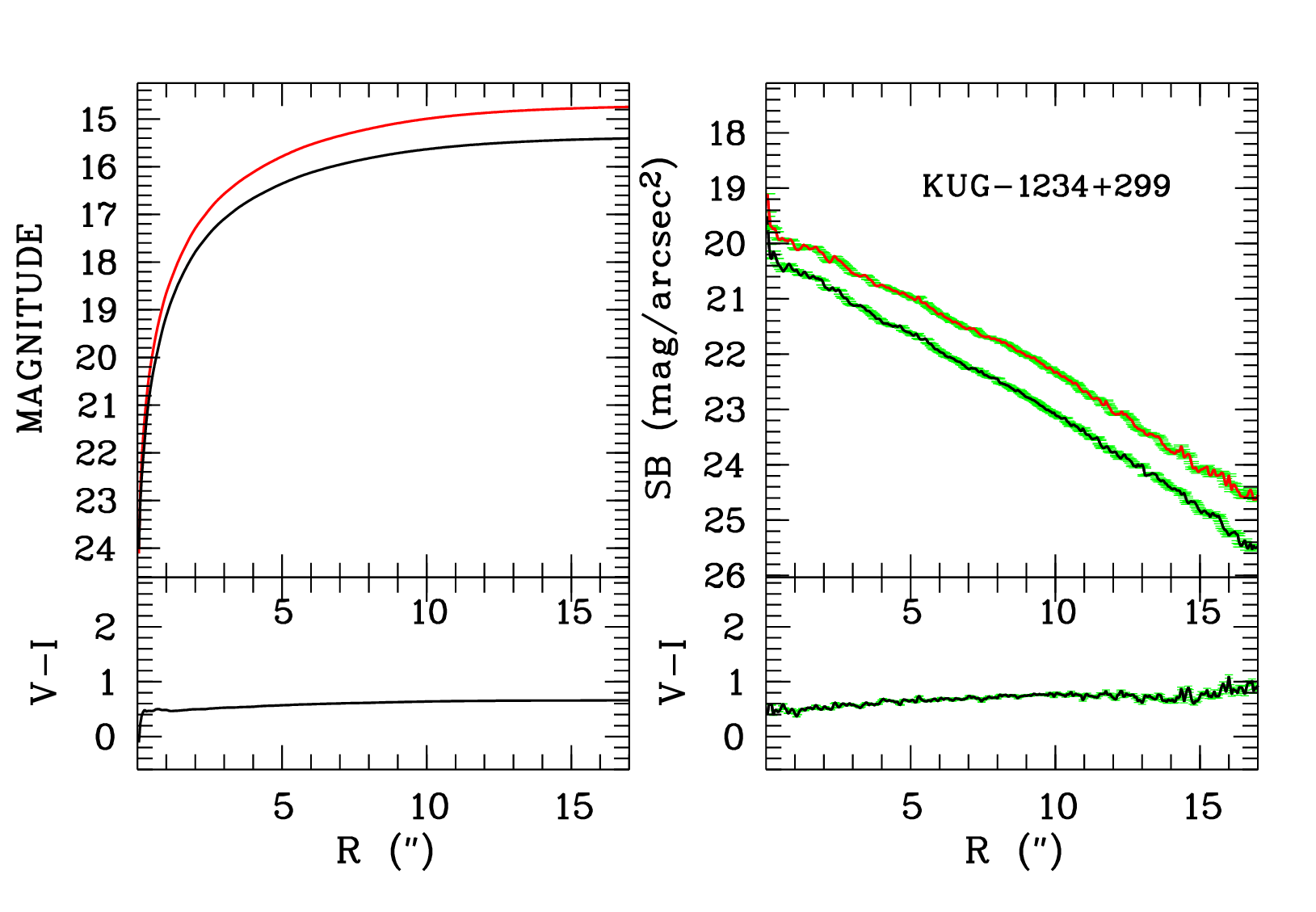}
\end{minipage}
\hfill
\begin{minipage}[h]{0.47\linewidth}
\includegraphics[scale=0.27,angle=-90]{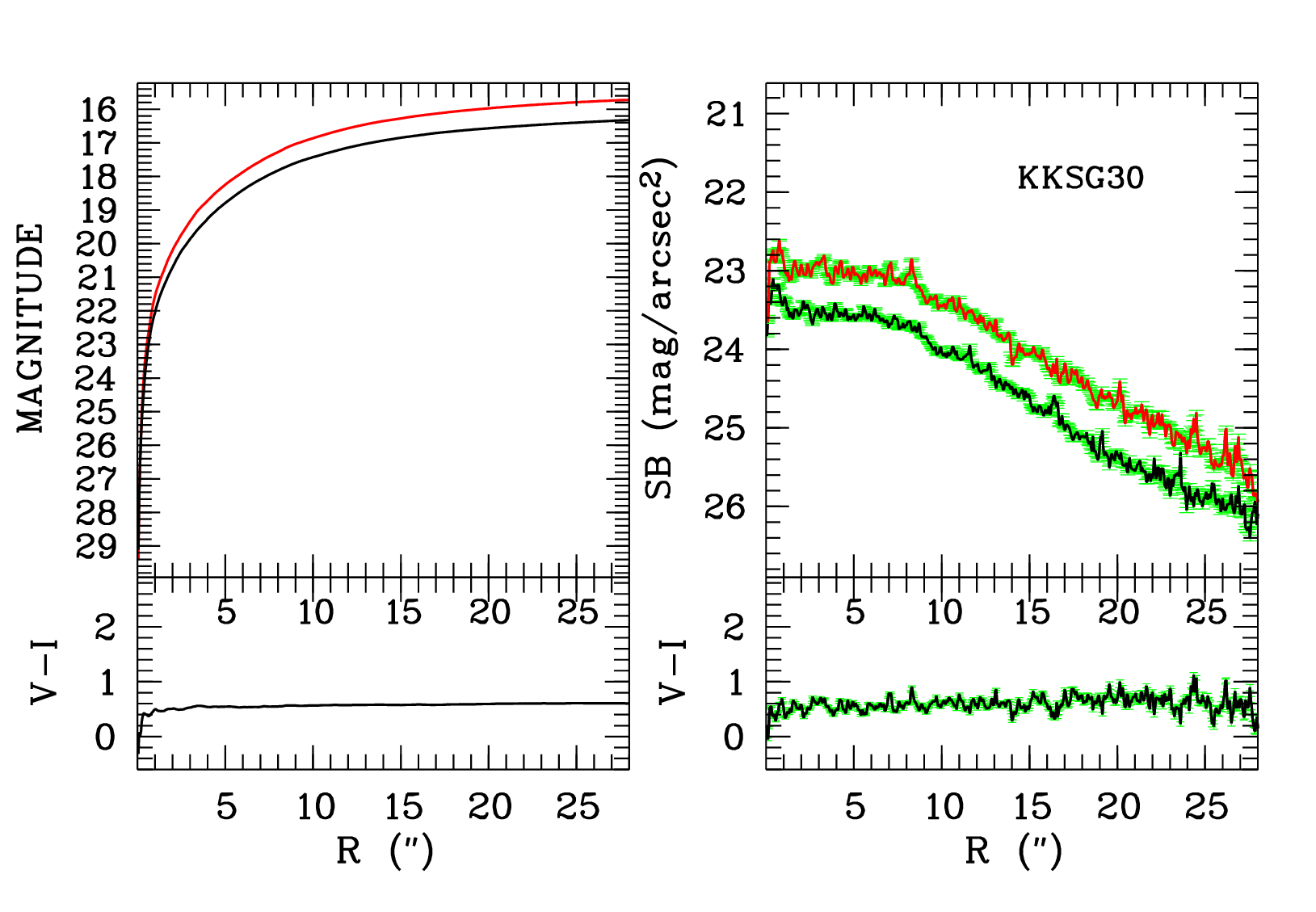}
\end{minipage}
\vfill
\begin{minipage}[h]{0.47\linewidth}
\includegraphics[scale=0.27,angle=-90]{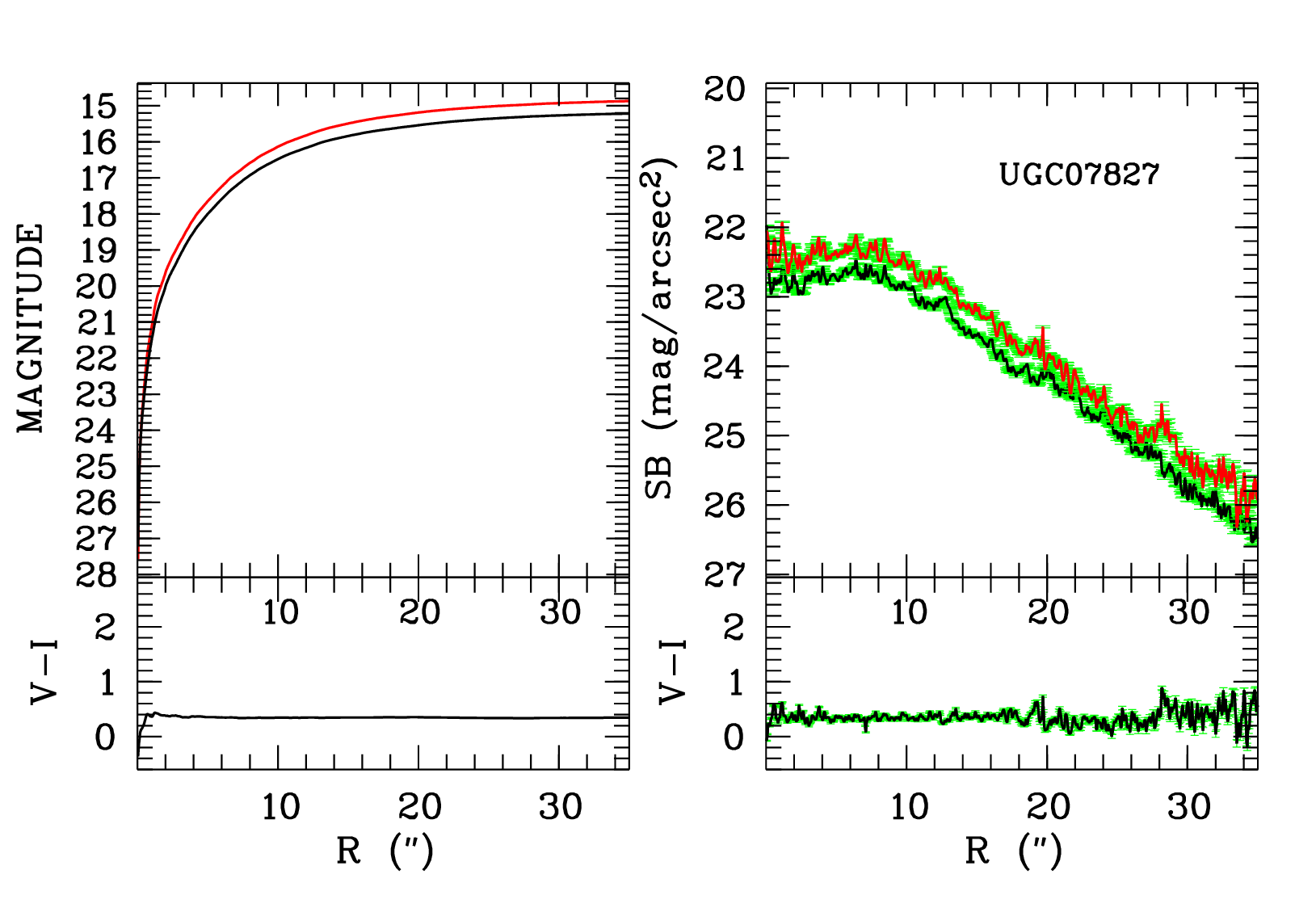}
\end{minipage}
\hfill
\begin{minipage}[h]{0.47\linewidth}
\includegraphics[scale=0.27,angle=-90]{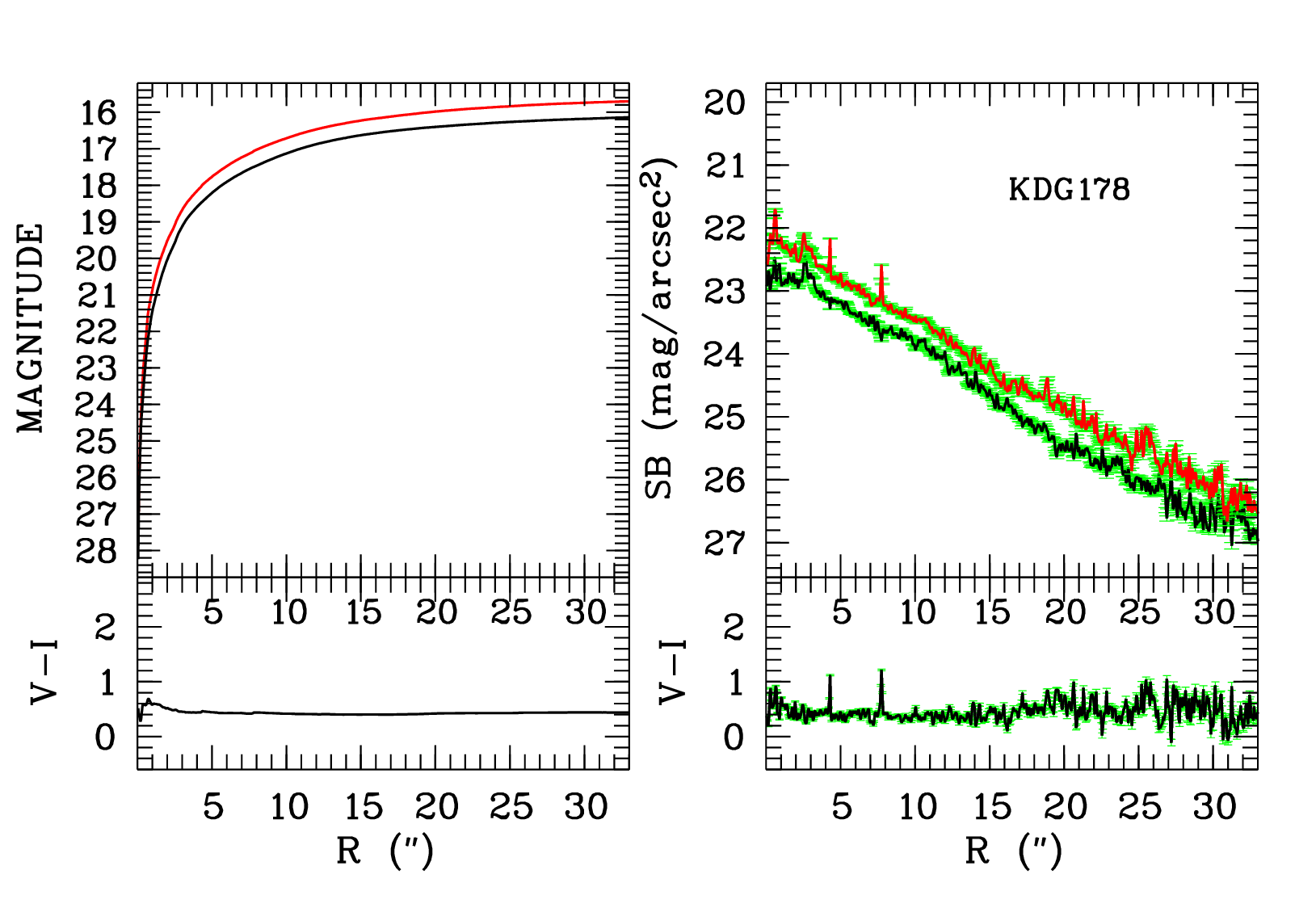}
\end{minipage}
\caption{Continued.}
\end{figure*}

\setcounter{figure}{0}
\begin{figure*}
\begin{minipage}[h]{0.47\linewidth}
\includegraphics[scale=0.27,angle=-90]{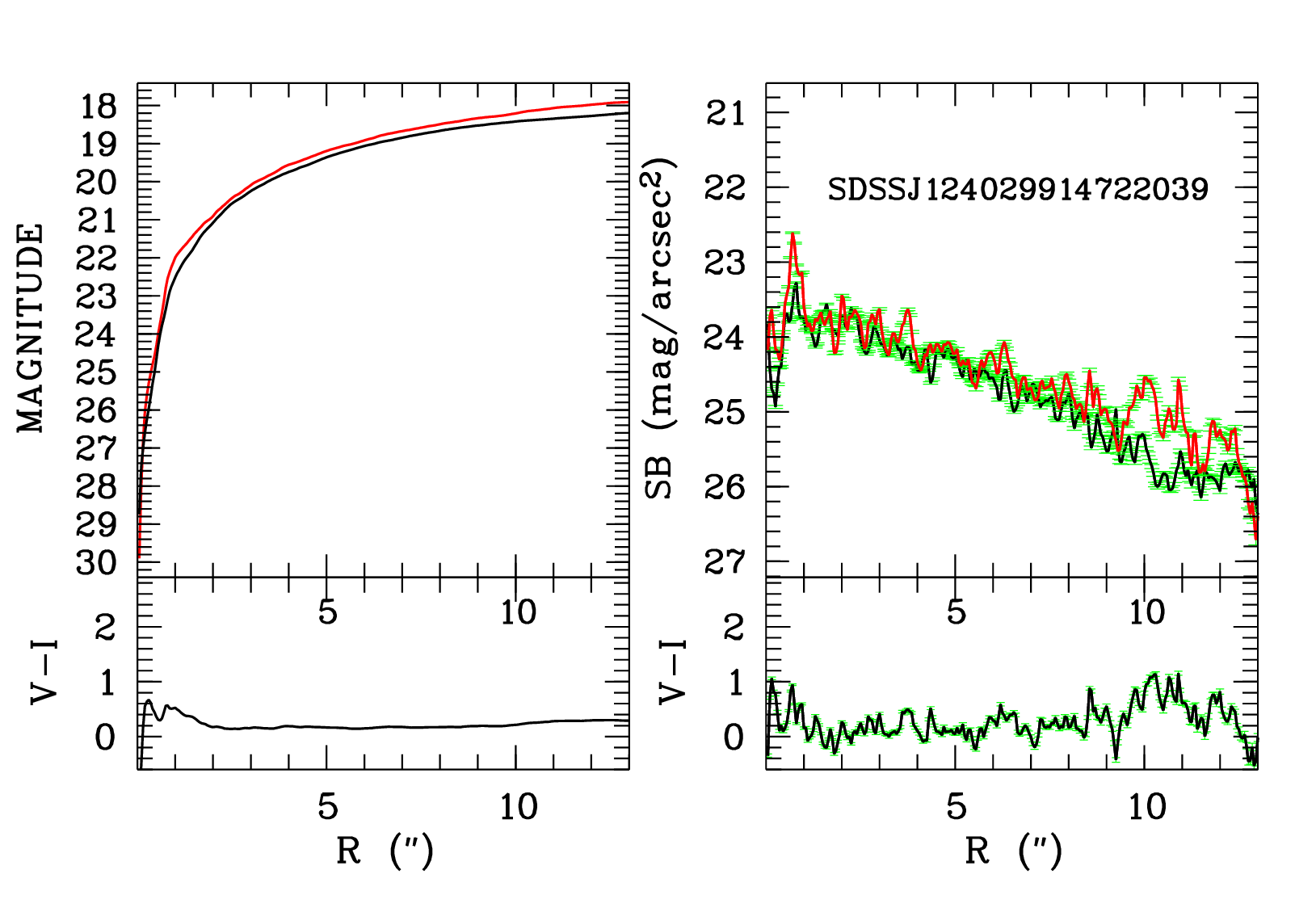}
\end{minipage}
\hfill
\begin{minipage}[h]{0.47\linewidth}
\includegraphics[scale=0.27,angle=-90]{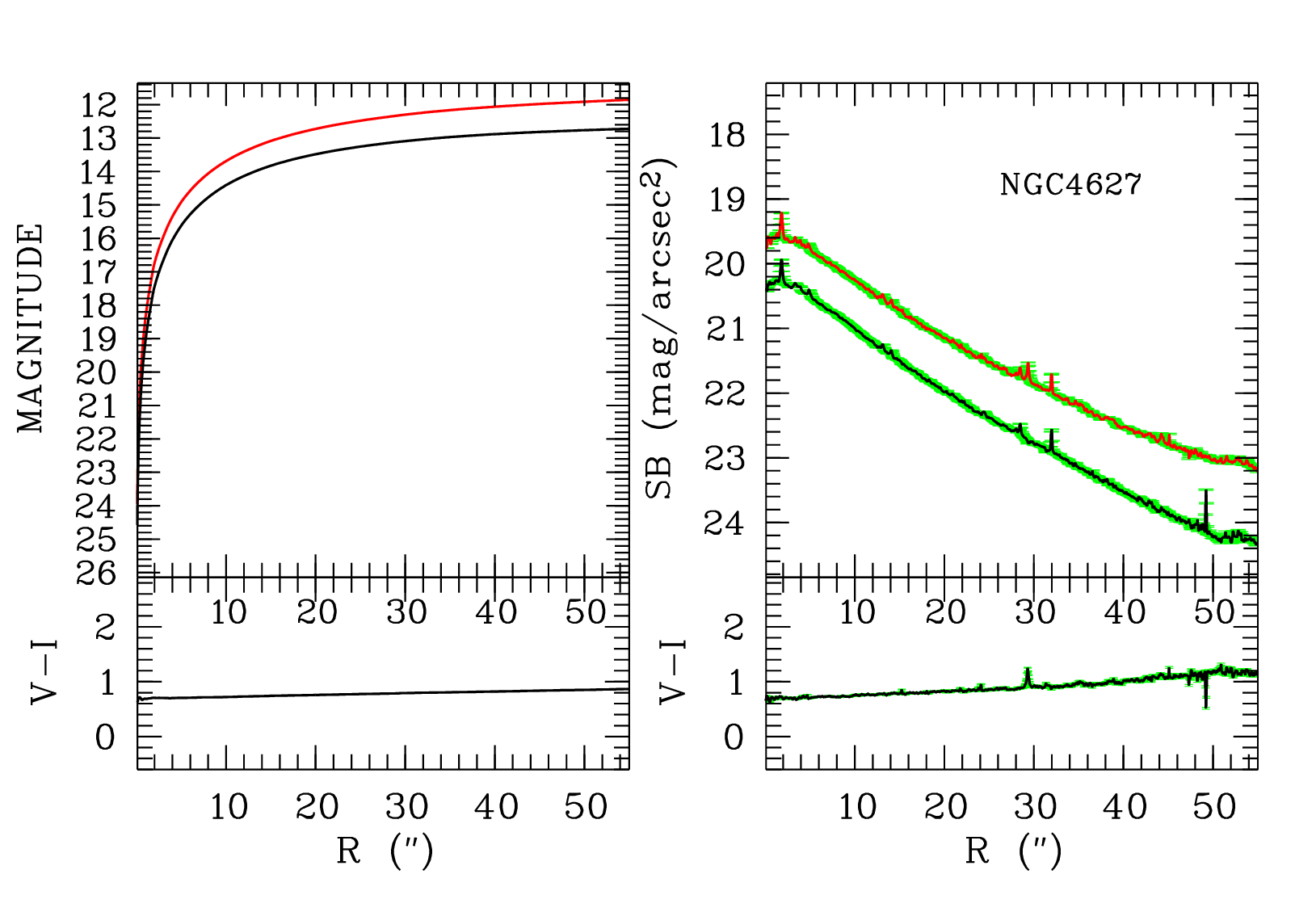}
\end{minipage}
\vfill
\begin{minipage}[h]{0.47\linewidth}
\includegraphics[scale=0.27,angle=-90]{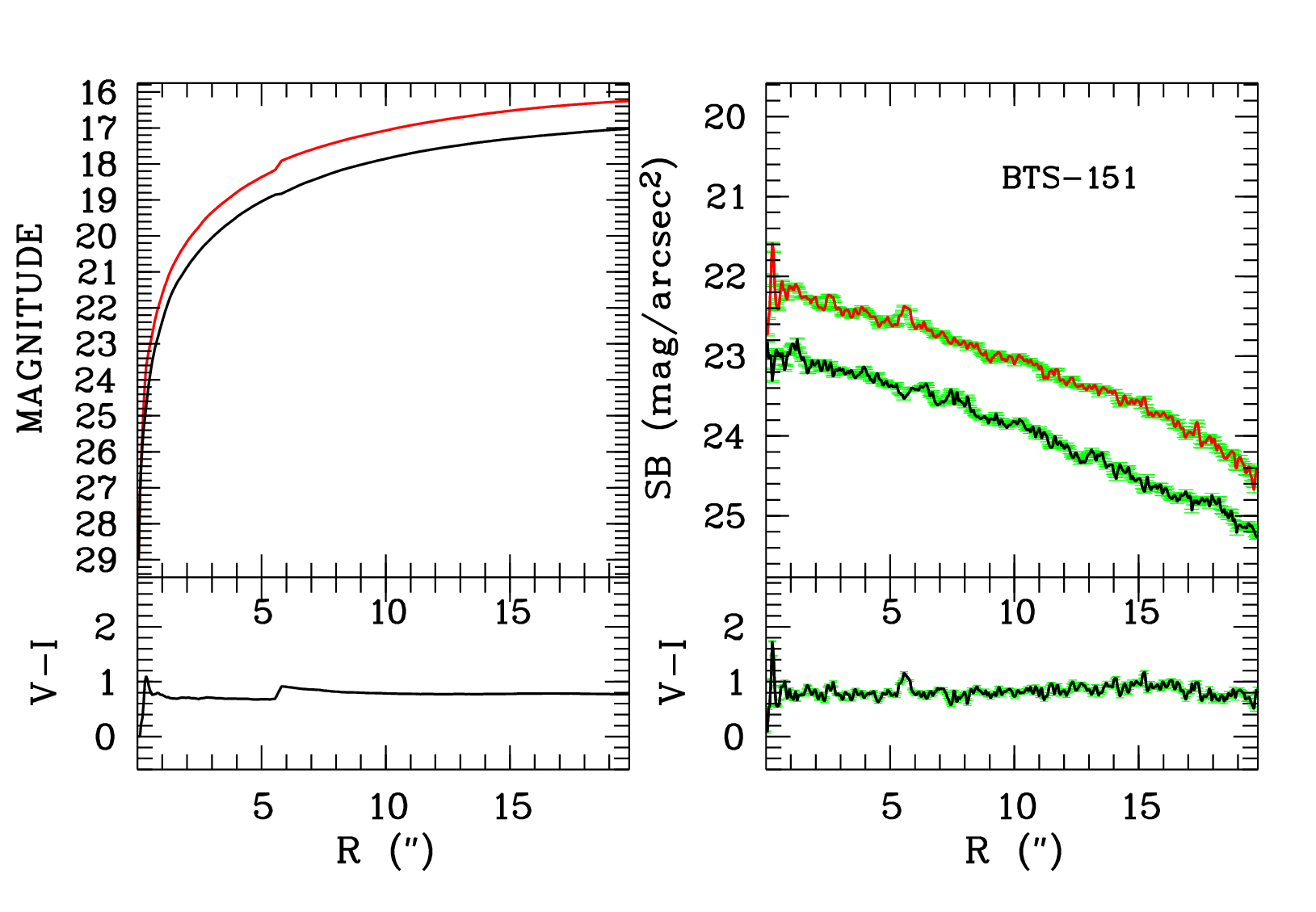}
\end{minipage}
\hfill
\begin{minipage}[h]{0.47\linewidth}
\includegraphics[scale=0.27,angle=-90]{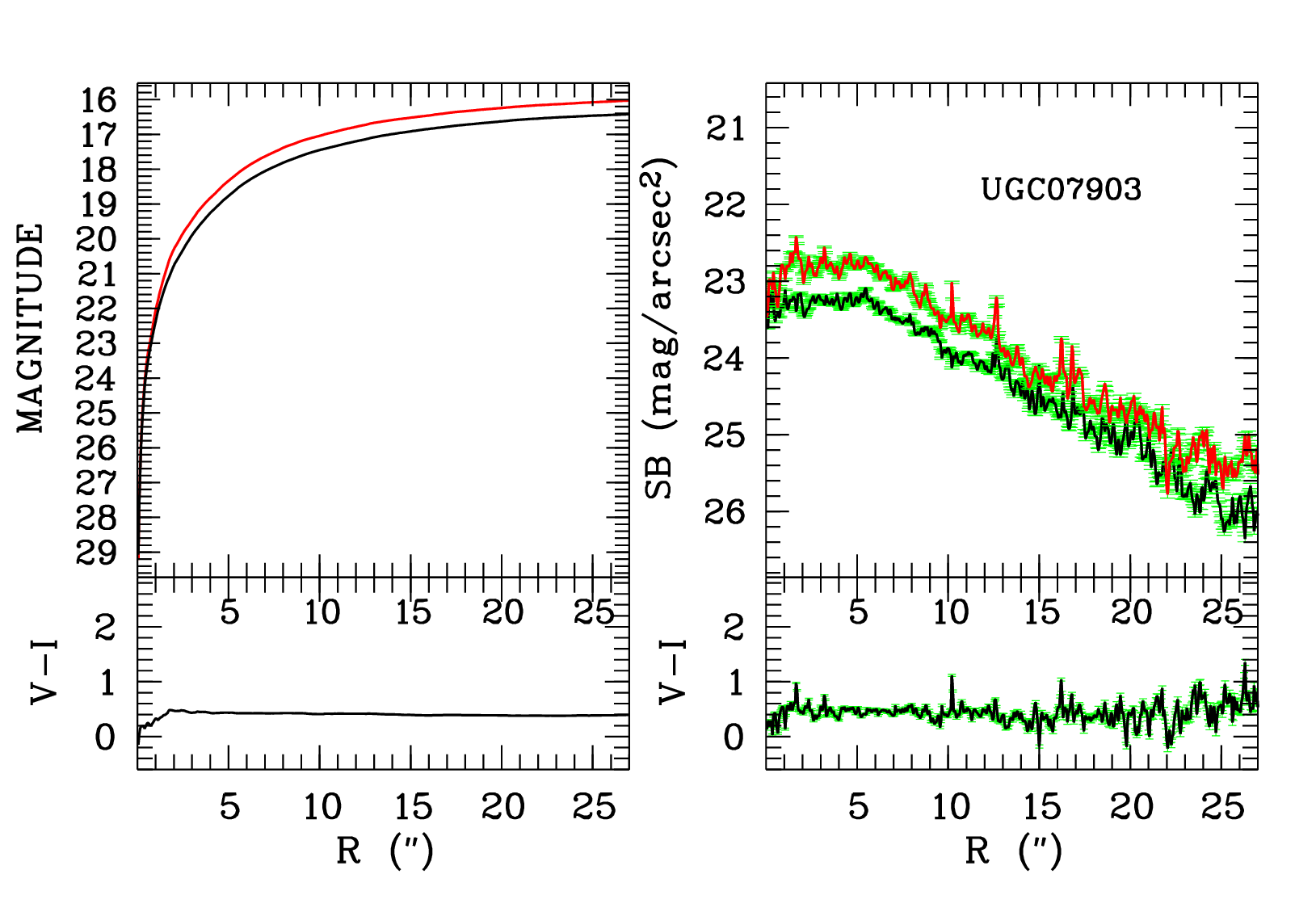}
\end{minipage}
\vfill
\begin{minipage}[h]{0.47\linewidth}
\includegraphics[scale=0.27,angle=-90]{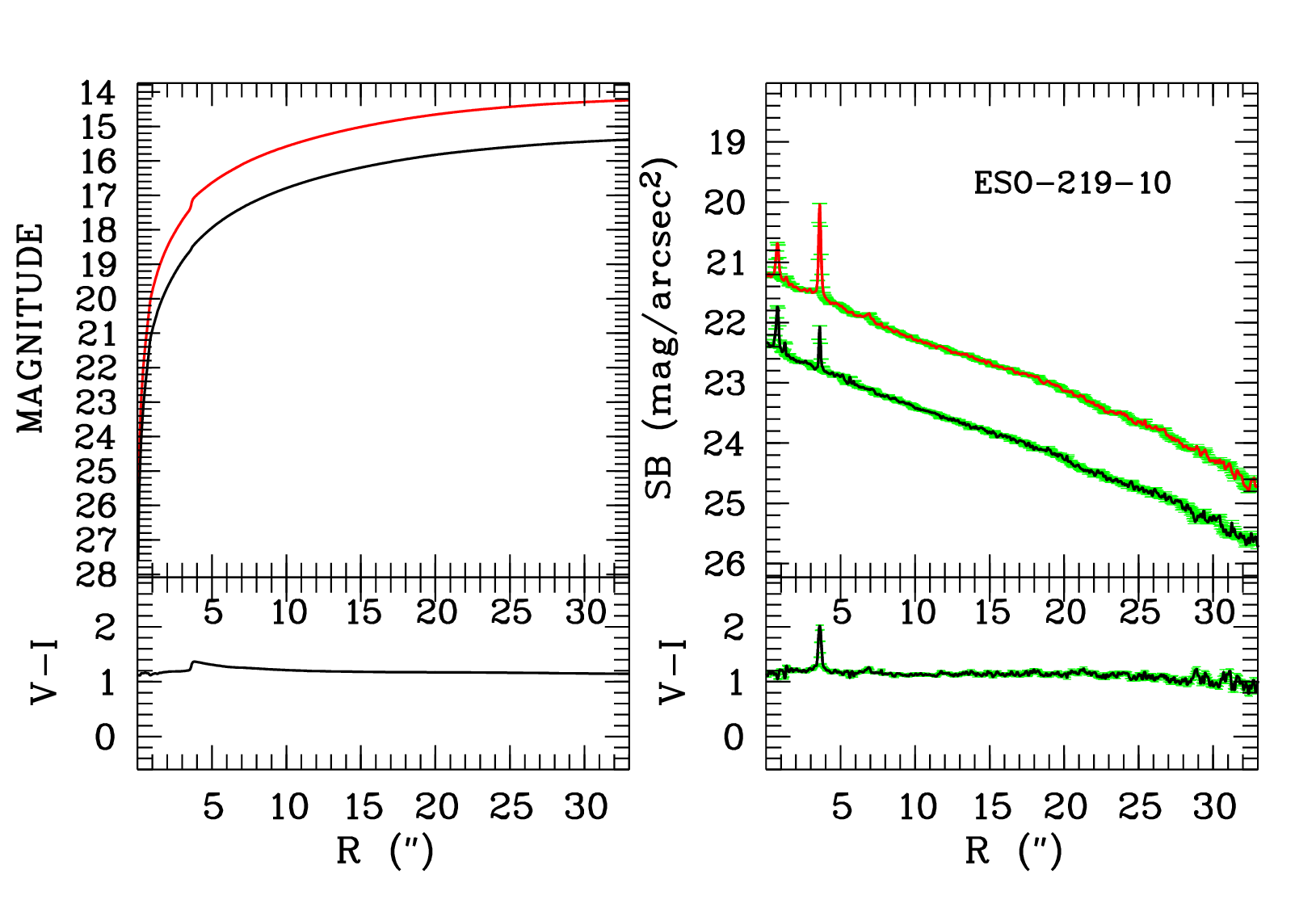}
\end{minipage}
\hfill
\begin{minipage}[h]{0.47\linewidth}
\includegraphics[scale=0.27,angle=-90]{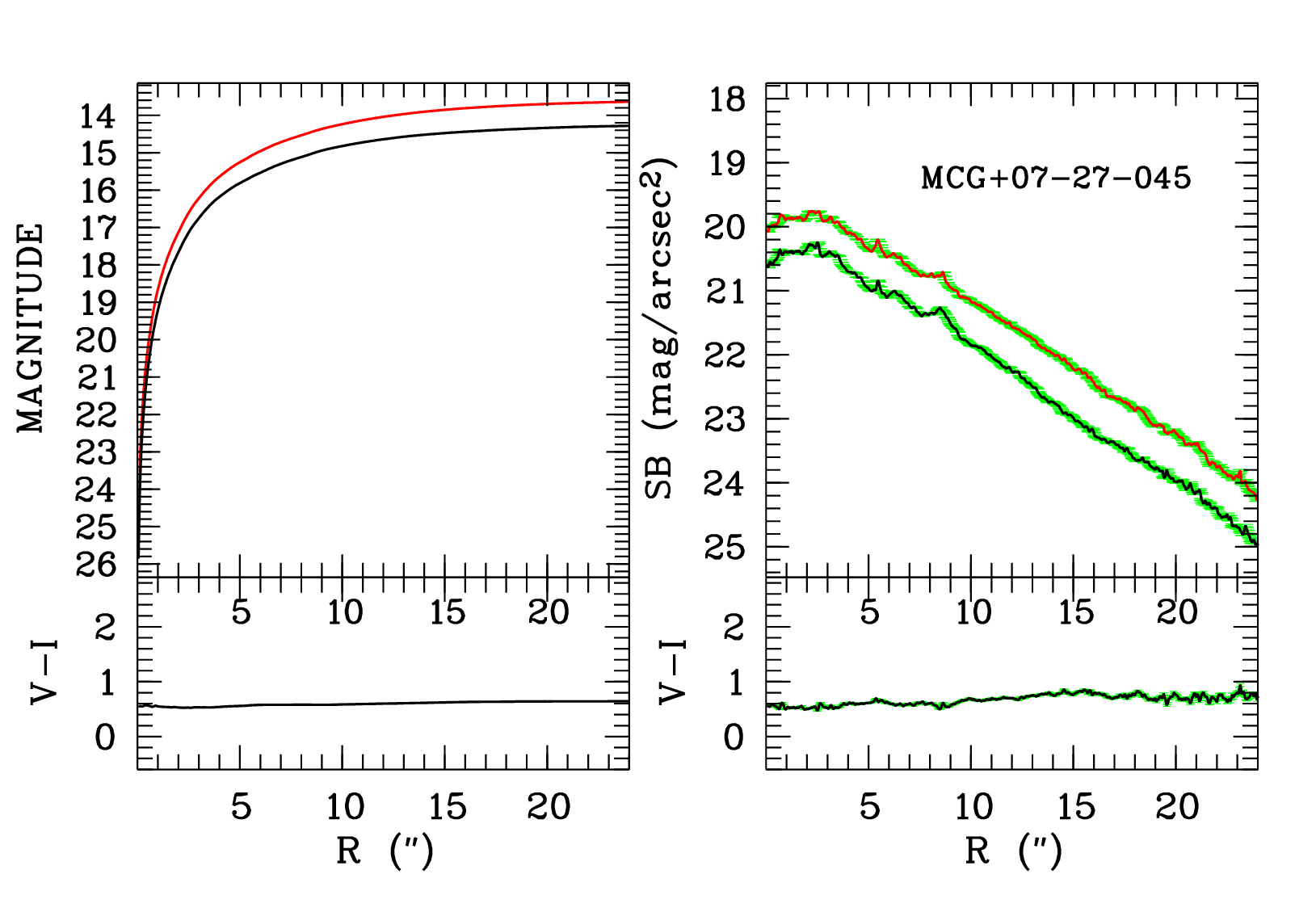}
\end{minipage}
\vfill
\begin{minipage}[h]{0.47\linewidth}
\includegraphics[scale=0.27,angle=-90]{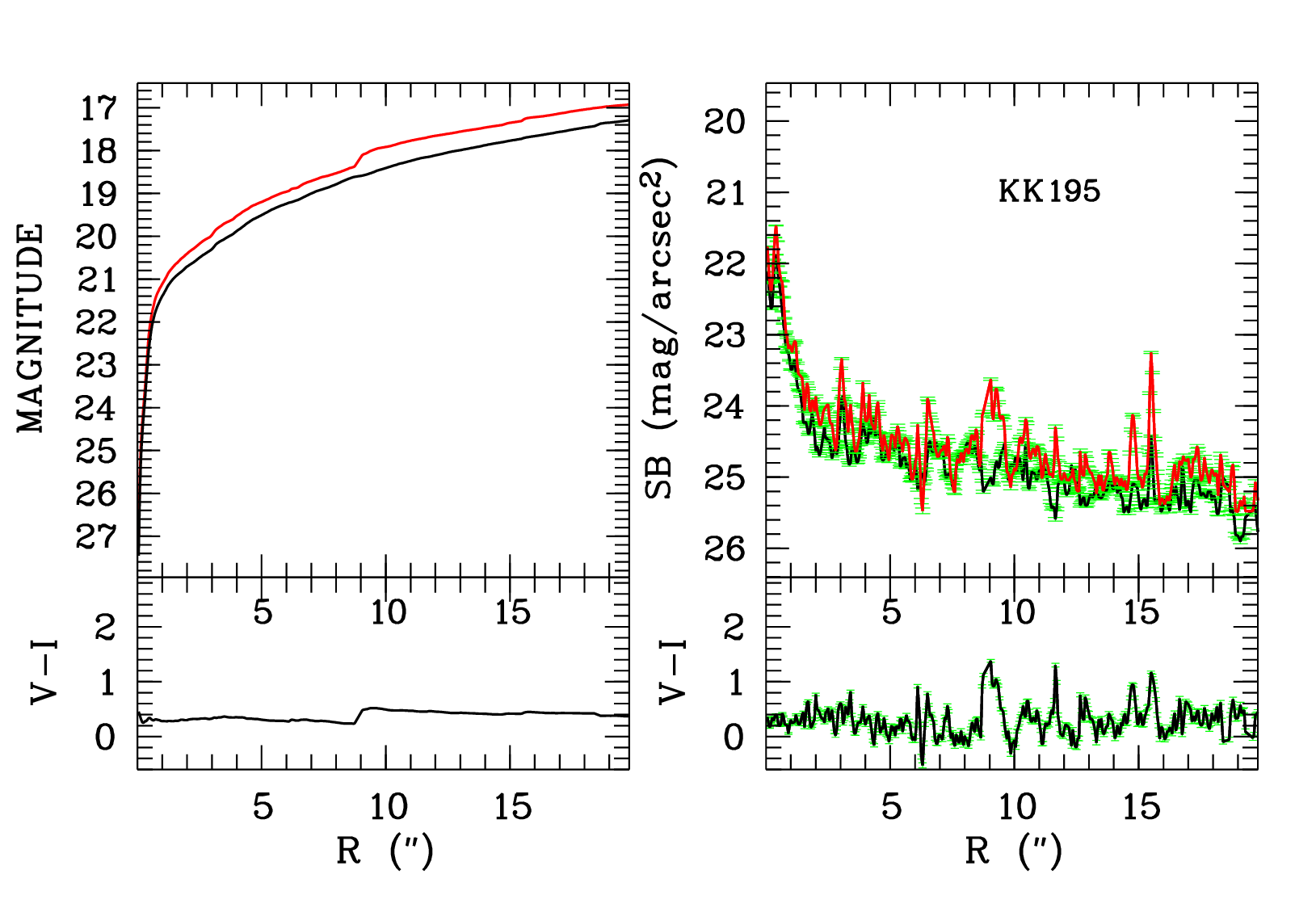}
\end{minipage}
\hfill
\begin{minipage}[h]{0.47\linewidth}
\includegraphics[scale=0.27,angle=-90]{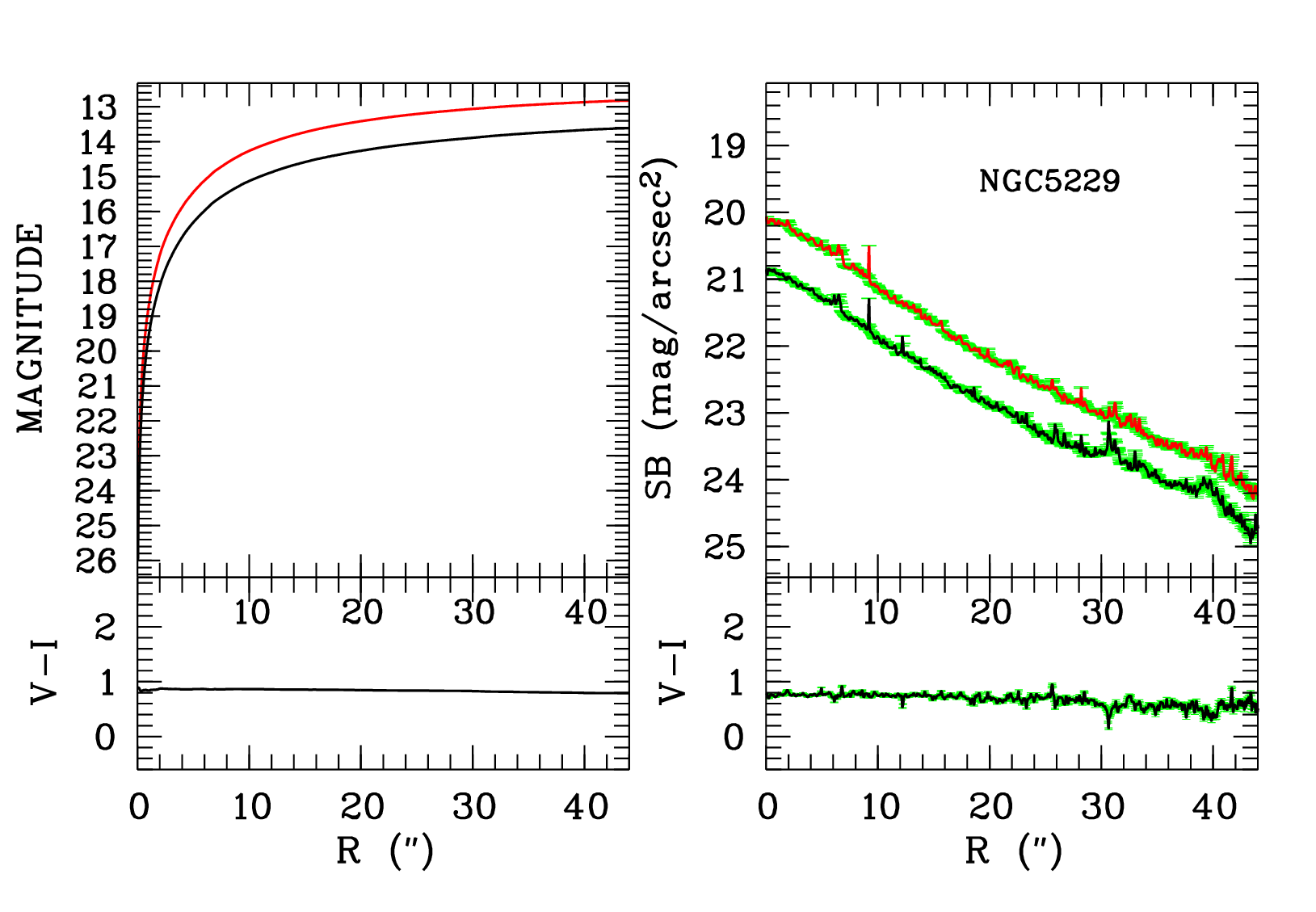}
\end{minipage}
\caption{Continued.}
\end{figure*}

\setcounter{figure}{0}
\begin{figure*}
\begin{minipage}[h]{0.47\linewidth}
\includegraphics[scale=0.27,angle=-90]{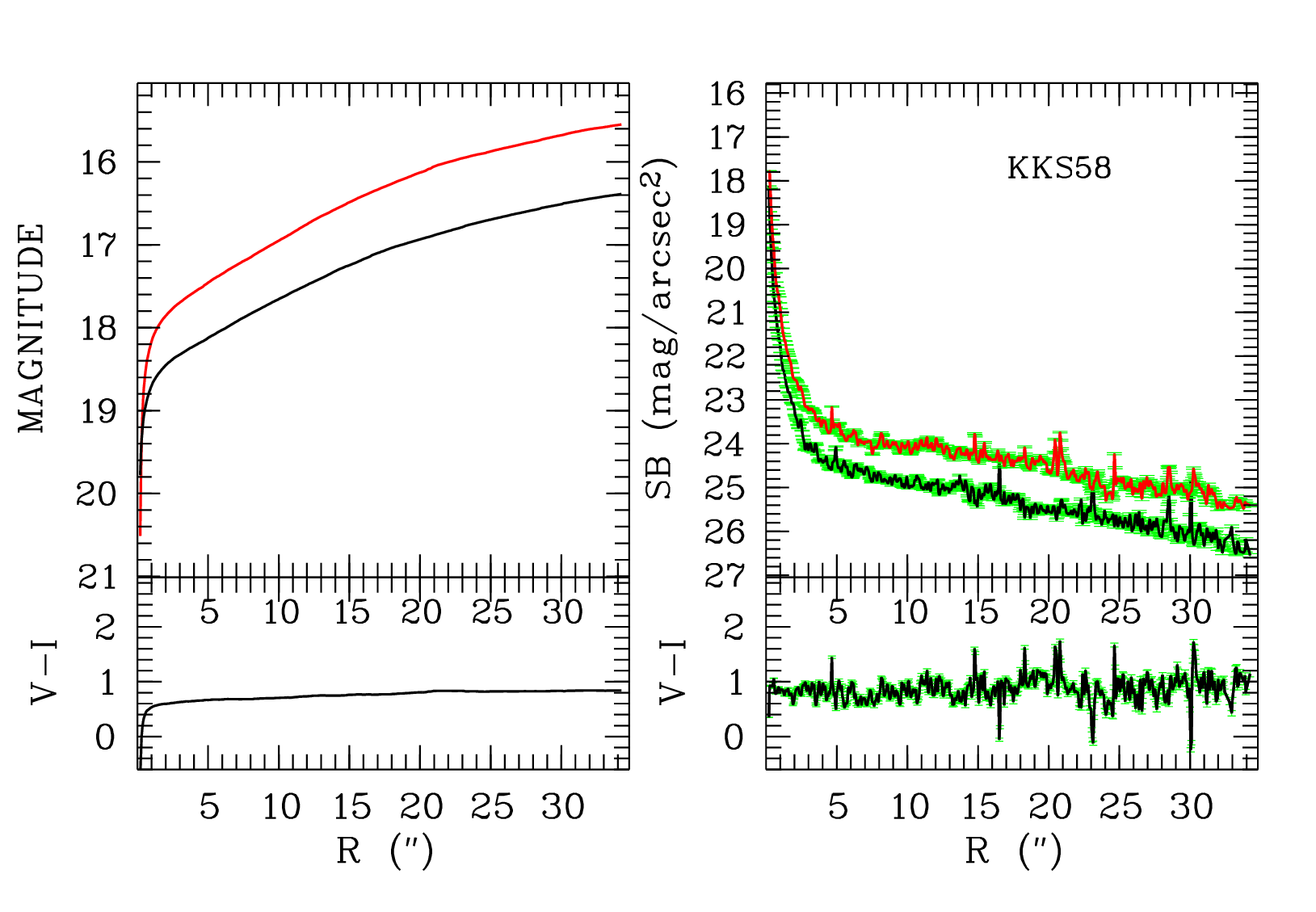}
\end{minipage}
\hfill
\begin{minipage}[h]{0.47\linewidth}
\includegraphics[scale=0.27,angle=-90]{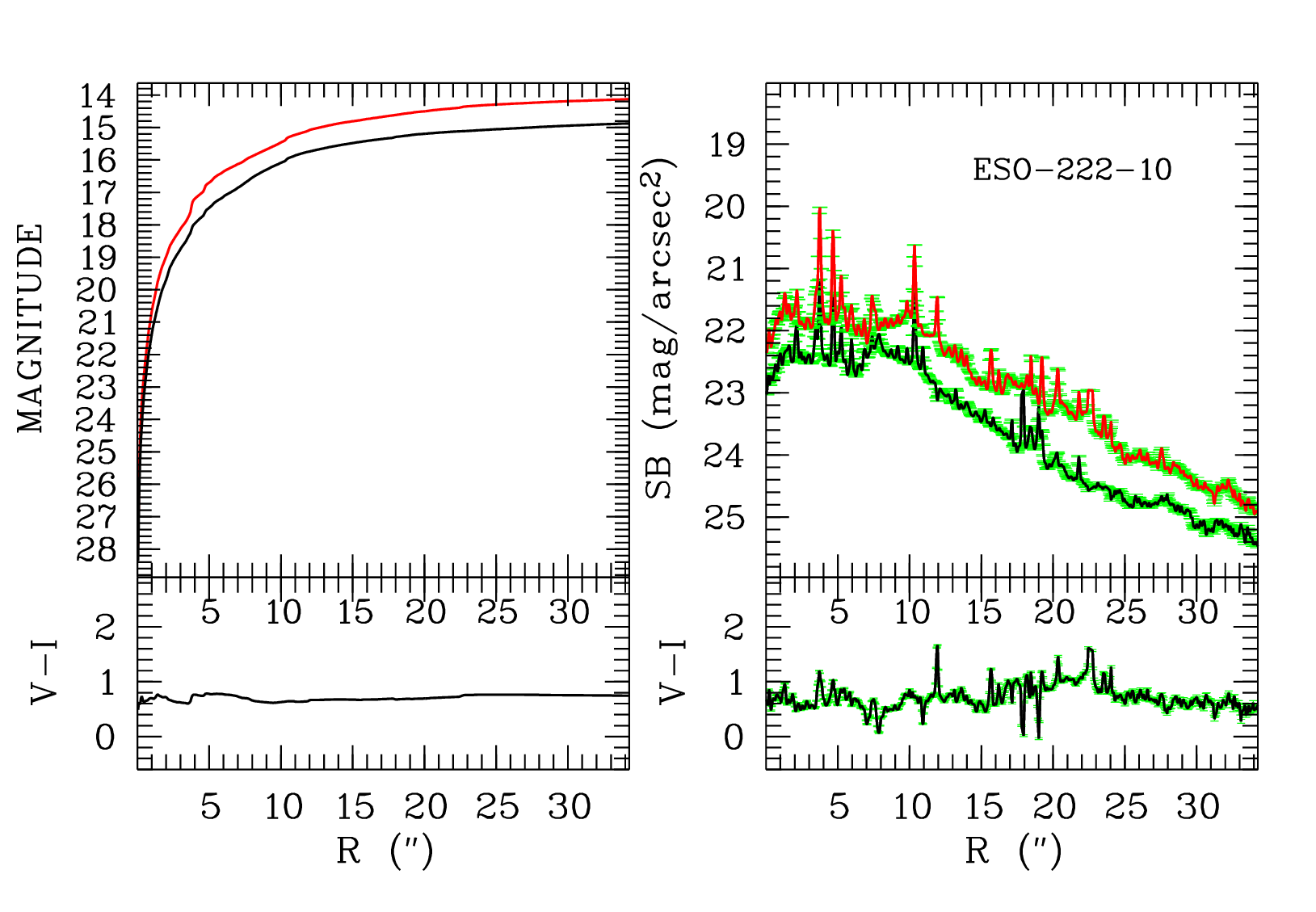}
\end{minipage}
\vfill
\begin{minipage}[h]{0.47\linewidth}
\includegraphics[scale=0.27,angle=-90]{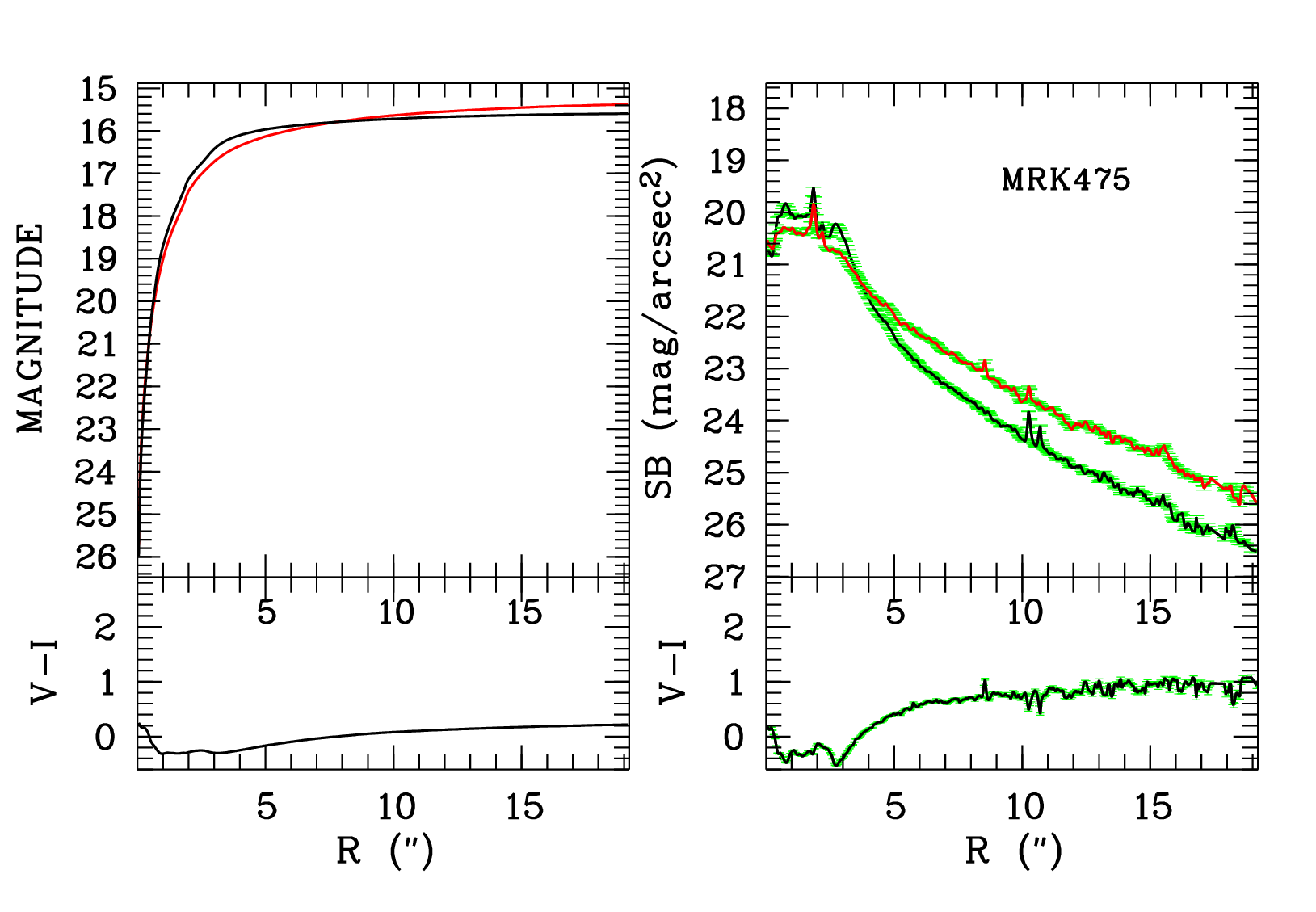}
\end{minipage}
\hfill
\begin{minipage}[h]{0.47\linewidth}
\includegraphics[scale=0.27,angle=-90]{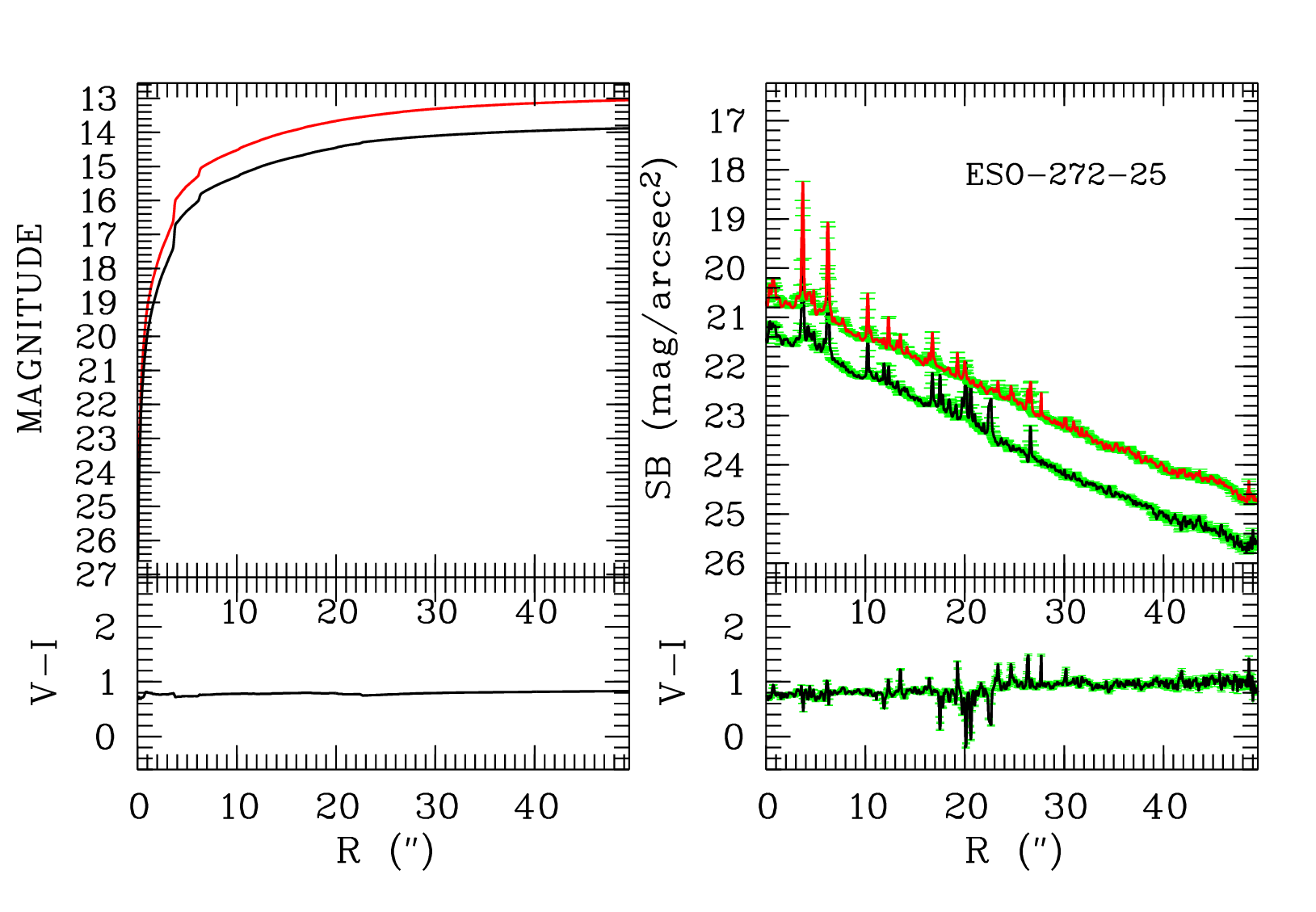}
\end{minipage}
\vfill
\begin{minipage}[h]{0.47\linewidth}
\includegraphics[scale=0.27,angle=-90]{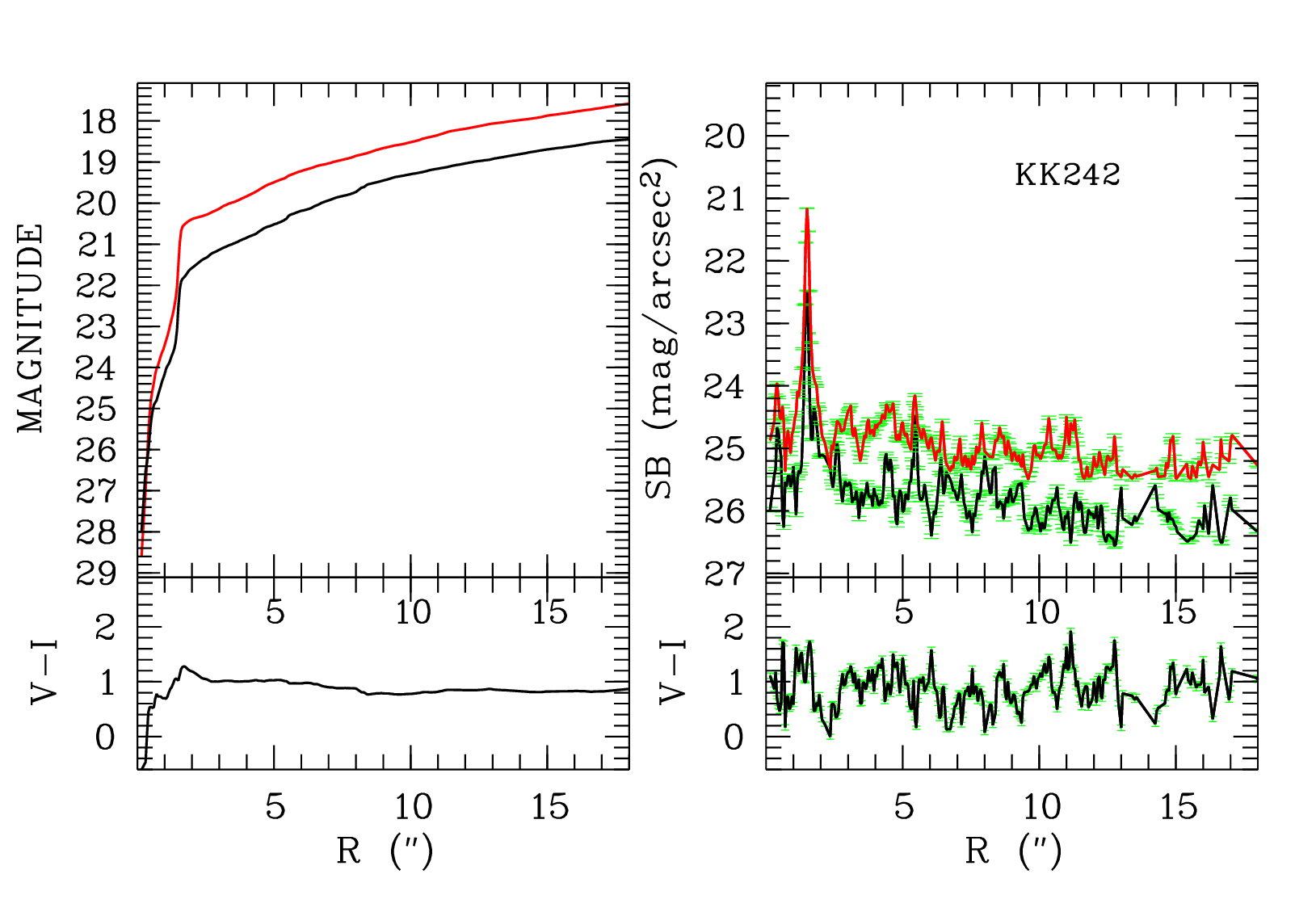}
\end{minipage}
\hfill
\begin{minipage}[h]{0.47\linewidth}
\includegraphics[scale=0.27,angle=-90]{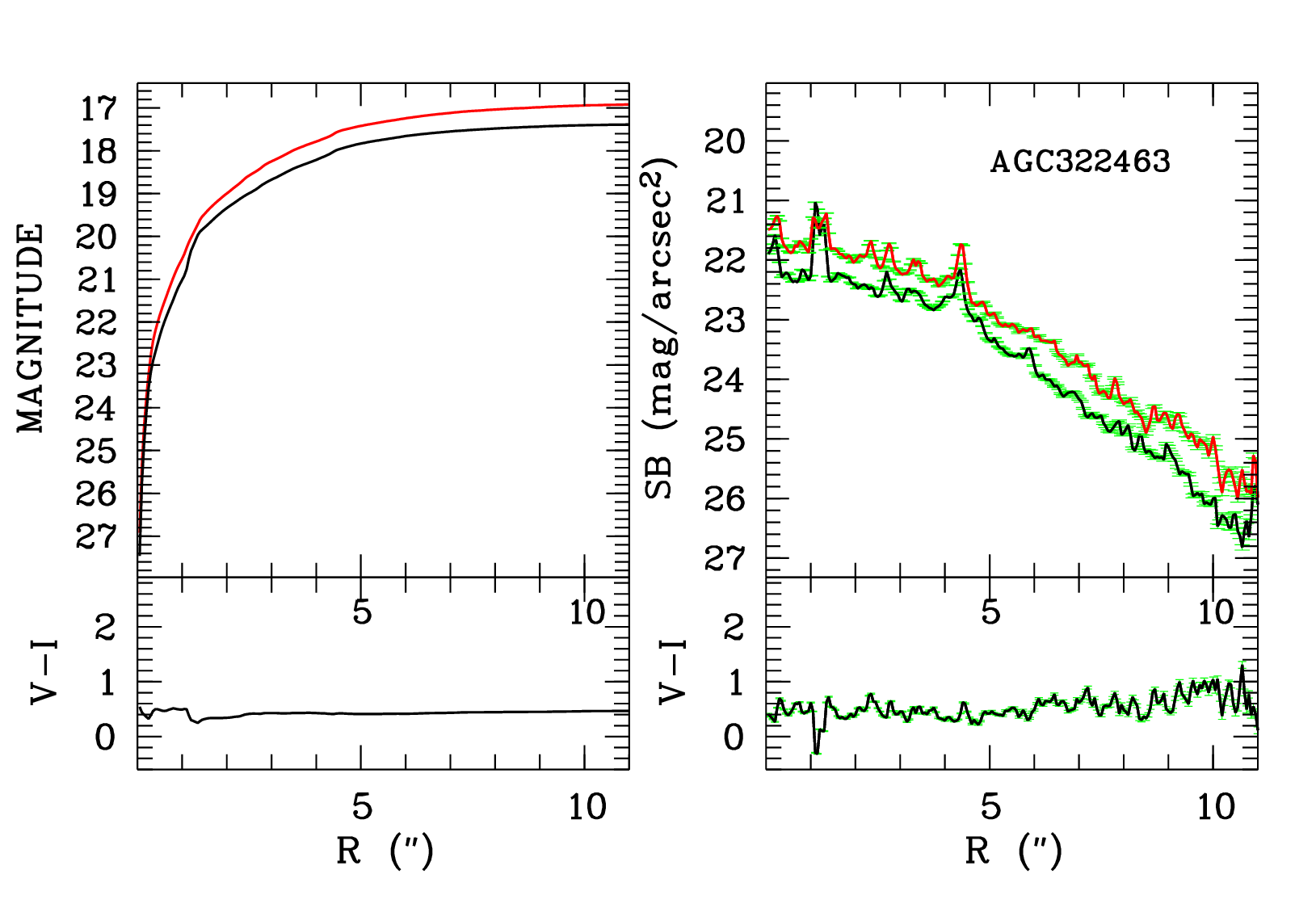}
\end{minipage}
\vfill
\begin{minipage}[h]{0.47\linewidth}
\includegraphics[scale=0.27,angle=-90]{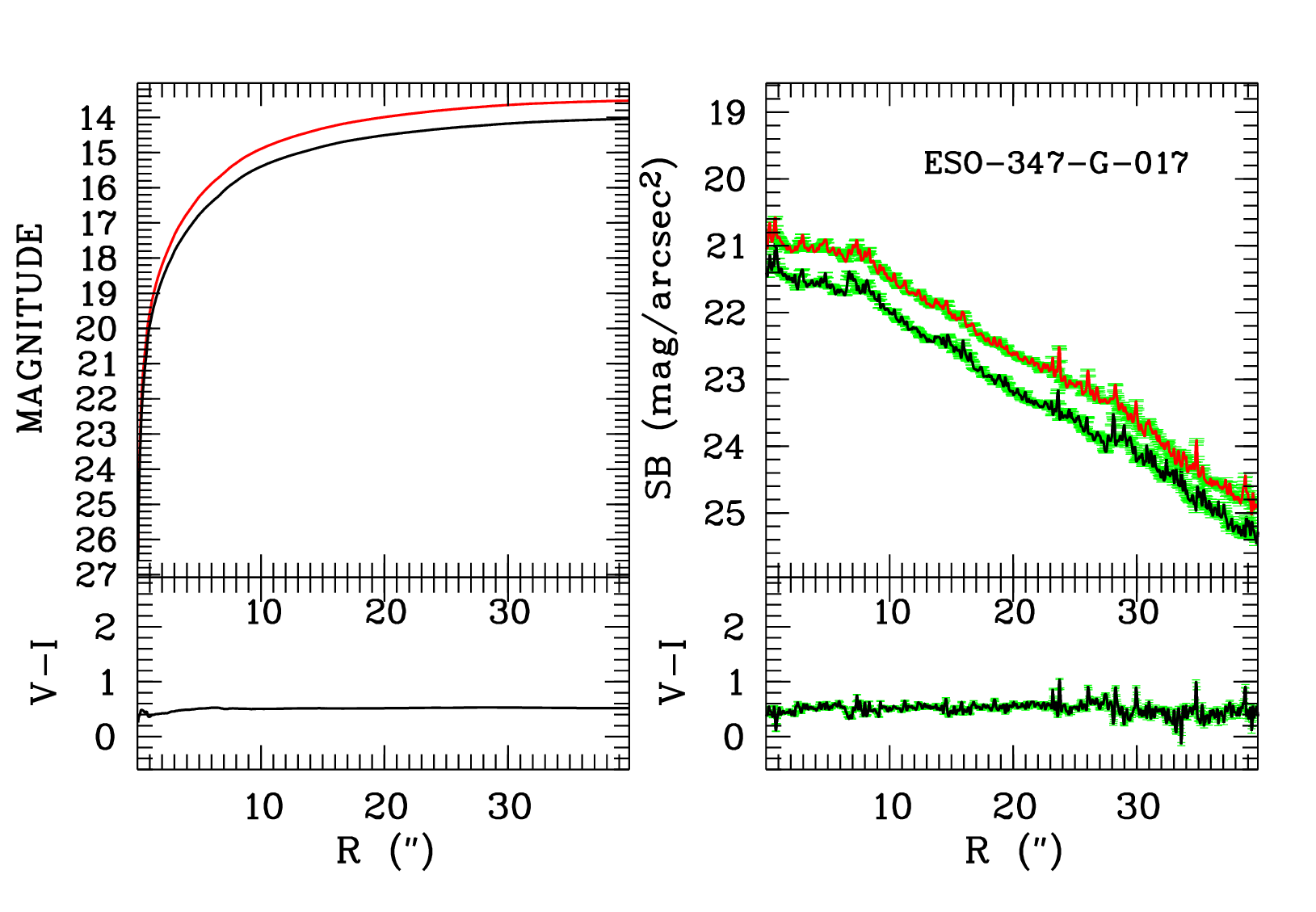}
\end{minipage}
\hfill
\begin{minipage}[h]{0.47\linewidth}
\includegraphics[scale=0.27,angle=-90]{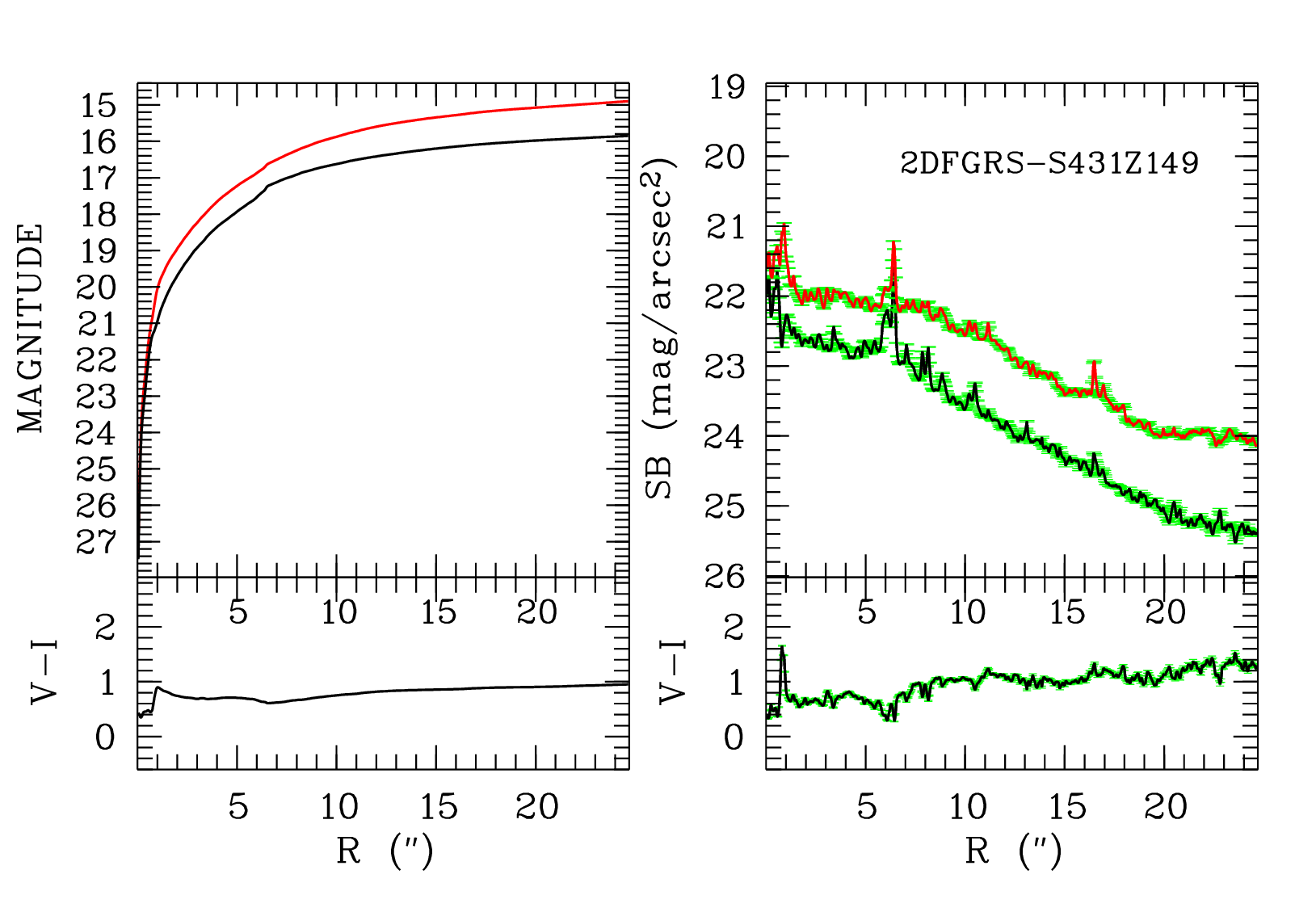}
\end{minipage}
\caption{Continued.}
\end{figure*}

\end{document}